\documentclass{article}
\usepackage[utf8]{inputenc}

\title{How to learn from inconsistencies: Integrating molecular simulations with experimental data}
\author{Simone Orioli$^{1,2}$*, Andreas Haahr Larsen$^{1,2,3}$*, Sandro Bottaro$^{1,4}$\\
and Kresten Lindorff-Larsen$^1$}

\usepackage{graphicx}
\usepackage{adjustbox}
\usepackage[numbers,sort&compress]{natbib}
\usepackage{amsmath}
\usepackage{enumitem}
\begin{document}

\date{}%
\maketitle

\noindent
{\small
$^1$ Structural Biology and NMR Laboratory \& Linderstrøm-Lang Centre for Protein Science, Department of Biology, University of
Copenhagen, Copenhagen, Denmark. %
$^2$ Structural Biophysics, Niels Bohr Institute, Faculty of Science, University of Copenhagen, Copenhagen, Denmark. %
$^3$ Present address: Structural Bioinformatics and Computational Biochemistry Unit, Department of Biochemistry, University of Oxford, Oxford, United Kingdom. %
$^4$ Atomistic Simulations Laboratory, Istituto Italiano di Tecnologia, Genova,
Italy. $^*$ These authors contributed equally to this work\\[1ex]
}
%}
\enlargethispage{\baselineskip}
\begin{abstract}
Molecular simulations and biophysical experiments can be used to provide independent and complementary insights into the molecular origin of biological processes. A particularly useful strategy is to use molecular simulations as a modelling tool to interpret experimental measurements, and to use experimental data to refine our biophysical models. Thus, explicit integration and synergy between molecular simulations and experiments is fundamental for furthering our understanding of biological processes. This is especially true in the case where discrepancies between measured and simulated observables emerge. In this chapter, we provide an overview of some of the core ideas behind methods that were developed to improve the consistency between experimental information and numerical predictions. We distinguish between situations where experiments are used to refine our understanding and models of specific systems, and situations where experiments are used more generally to refine transferable models. We discuss different philosophies and attempt to unify them in a single framework. Until now, such integration between experiments and simulations have mostly been applied to equilibrium data, and we discuss more recent developments aimed to analyse time-dependent or time-resolved data.
\end{abstract}

\newpage
\enlargethispage{\baselineskip}
\tableofcontents
\newpage

\section{Introduction}\label{sec:introduction}
Molecular mechanics simulations of biological systems have matured over the last decades, to a state where reliable predictions and interpretations of biological phenomena can be achieved \cite{Lopes2015,bottaro2018biophysical}. Simulations can readily be used either before experiments to suggest hypotheses and design experiments, or after experiments to analyse and interpret the data. With simulations it is possible to probe timescales and spatial details that are, yet, impossible to access experimentally, and thus they provide a unique tool to study, e.g. conformational transitions \cite{Maximova2016,Adcock2006,Klepeis2009,Abrams2013,elber2016perspective}, signalling events \cite{hollingsworth2018molecular}, ion transport \cite{hollingsworth2018molecular,maffeo2012modeling}, and many other phenomena.
Since most experimental observables are averaged over time and a large number of molecules, simulations can help resolve the underlying structural and dynamical distribution with atomistic spatial accuracy and femtosecond time resolution. 

As simulations are in essence a theoretical framework, experimental verification is pivotal and there are still, in many cases, significant differences between experimental data and the corresponding observables calculated from state-of-the art simulations \cite{van2018validation}. These discrepancies are usually due to imperfect force fields \cite{rauscher2015structural,henriques2015molecular,robustelli2018developing,nerenberg2018new}, insufficient sampling \cite{bernardi2015enhanced} or inaccurate forward models \cite{Cordeiro2017}. Even with accurate forward models and robust sampling, however, the quality of the simulation results are bound by the accuracy of the underlying force field \cite{robustelli2018developing}. Indeed, molecular mechanics is intrinsically limited by the fact that it employs classical approximations for phenomena that are known to happen on a quantum scale. Force fields have been developed to approximate these quantum interactions within a classical framework, in such a way to balance between computational simplicity and accuracy: thus, an exact match is not to be expected and synergy between simulations and experiments remains of fundamental importance. Substantial efforts of the biophysical community have been therefore directed towards the development of methods that correct the results of simulations by accounting for experimental information, either in a system specific or generalisable fashion.  In this chapter we provide a self-contained overview of the principles and strategies that have been applied to infer conformational ensembles, i.e. collections of structures reflecting the dynamism and plasticity of a molecule, as well as to improve force fields using experimental data. In other words, we will make a clear cut distinction between system-specific and general force field corrections: in the former case, simulations are performed according to experimental conditions, and if inconsistencies emerge, the resulting ensemble is corrected according to the available experimental information; in the latter case, instead, discrepancies between simulations and experimental observables are used to actively improve the physical potential employed to describe the system. We note that the goal of the chapter is not to be comprehensive of all the literature that has been previously published: rather, it focuses on some cornerstone principles and methods that summarise the efforts of the community. For this reason we also focus mostly on the methods and theory, and only provide few examples of applications.

This chapter is meant for readers experienced in molecular simulations and comfortable with the fundamentals of statistics. It is particularly suited for scientists that approach the combination of experiments with simulations for the first time or readers interested in a non-technical overview of the field.

The chapter is comprised of seven sections, each one concerned with a different aspect of the literature on the subject. Section \ref{sec:reweighting} is dedicated to the description of some of the main strategies to obtain consistency between simulation and data by manipulating the ensemble \textit{after} simulations have been performed. Differently, in section \ref{sec:on-the-fly} we will discuss how consistency can be achieved by introducing a system-specific empirical energy term in the force field. In this case, the refinement step occurs \textit{before} the actual simulation, as the experimental bias will guide the sampling towards only the relevant regions of conformational space. In section \ref{sec:FF} we discuss how to employ experimental data to refine the physical description of macromolecules, i.e. the force field, instead of the conformational ensemble of a particular system. Like the biased force field approaches described in section \ref{sec:on-the-fly}, these methods also adjust the force field \textit{before} the production simulation. In section \ref{sec:time} we discuss how to combine time-dependent and time-resolved data with simulations. This is particularly relevant, as computational power \cite{piana2013atomic,lindorff2016picosecond,voelz2012slow,bowman2010atomistic} and enhanced sampling techniques \cite{Maximova2016,camilloni2018advanced} have pushed the reachable timescales to the ones resolved by some experimental techniques. In section \ref{sec:challenges} we discuss some of the most important overall challenges arising from combining experimental data and simulations, and, finally, in section \ref{sec:conclusions} we conclude by summarising similarities and differences between the various approaches. We thus aim to give an overview of the present state of the field as well as pointing out key aspects where further collective effort is needed to improve the predictive power of combined computational and experimental methods for structural studies of biological systems. 

\section{Reweighting strategies} \label{sec:reweighting}
\begin{figure}[tbp!]
\begin{center}
\includegraphics[width=\textwidth]{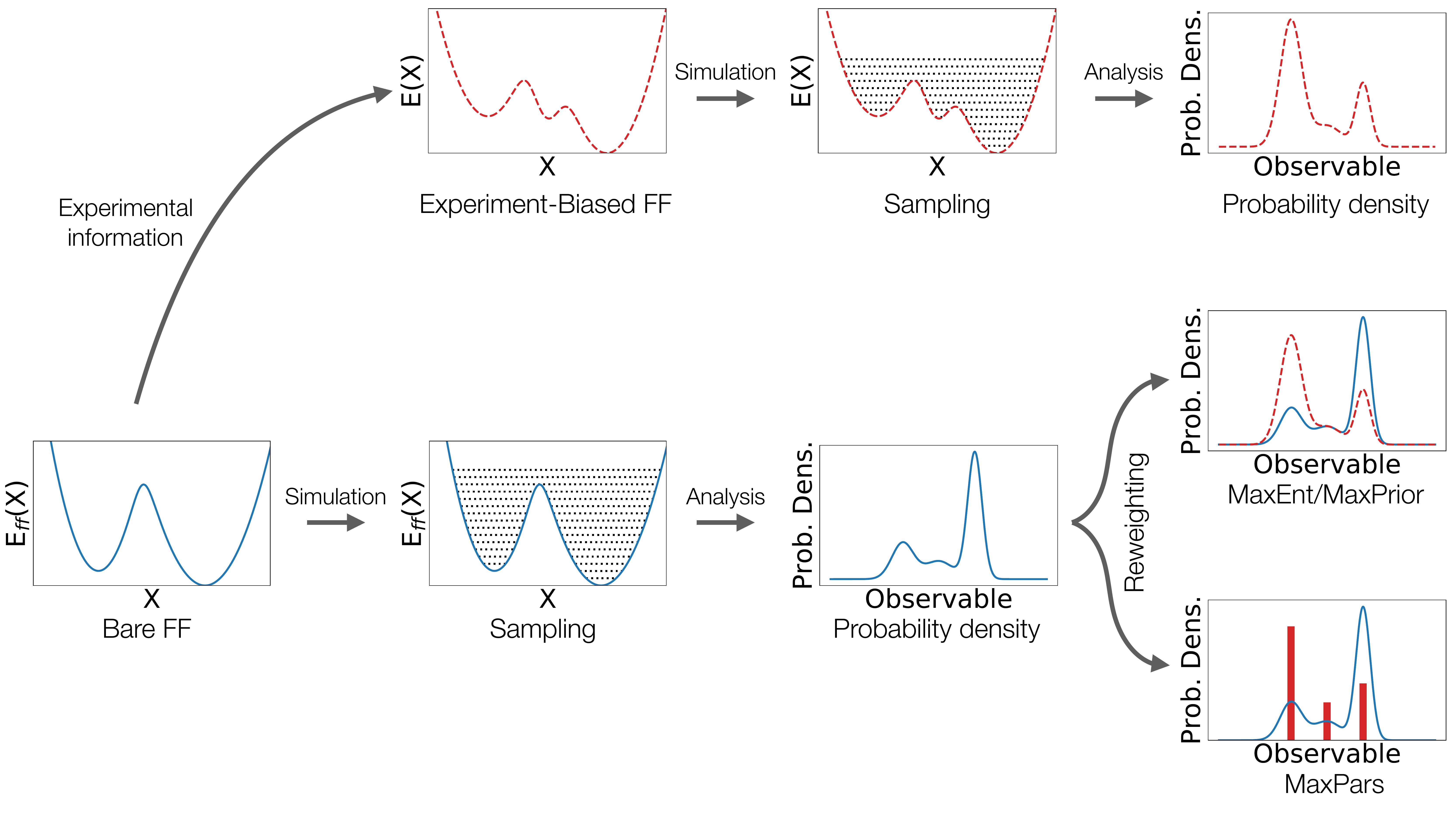}
\end{center}
\caption{Overview of methods to achieve consistency between simulation and experimental data for the distribution of a generic static observable. All methods start with an initial force field $E_\text{ff}(x)$ (`bare' FF, blue line) which is either employed as is or modified in accordance to some experimentally driven bias (experiment-biased FF, red dashed line). In the upper case, the biased force field is sampled to directly generate an observable distribution which is coherent with experimental data. In the lower case, the observable distribution is subsequently reweighted against experimental data using either the MaxEnt/MaxPrior or MaxPars principles to yield some reweighted distributions (red dashed line).}
\label{fig:overview}
\end{figure}

Consistency between simulations and experimental data can in many cases be achieved by reweighting a trajectory (or a set of trajectories) obtained with a given force field (Fig.~\ref{fig:overview}). Let us assume to perform a molecular dynamics (MD) or Monte Carlo (MC) simulation and to save $N$ snapshots, $X_i, \ldots, X_N$, from the trajectory for analysis. Each frame in the simulation is given an initial weight that reflects the population of that structure as predicted by the force field. We denote the initial set of weights, $\boldsymbol{w^0}= w^0_1,w^0_2,...,w^0_N$, the \emph{reference distribution} and we assume them to be non-negative and normalised as probabilities, $\sum_{i} w^0_i = 1$. When performing a standard MD/MC  simulation and under the assumption that enough sampling has been performed, the initial weights are constant, $w_i^0= 1/N$, as such simulations generate conformations distributed according to the Boltzmann distribution. When using enhanced sampling techniques that do not directly sample the Boltzmann distribution or, in the case of shorter off-equilibrium simulations, the initial weights are in general non-uniform \cite{wu2017variational}. Given the $N$ snapshots and the reference distribution, a static observable can, in the simplest case, be calculated as a weighted ensemble average:
\begin{align} \label{eq:ensemble_average}
    \langle O^\mathrm{calc}\rangle(\boldsymbol{w}) = \sum_{i=1}^N w_iO^\mathrm{calc}(X_i).
\end{align}
Some experimental quantities (e.g. NMR relaxation rates and other inherently time-dependent observables) cannot generally be expressed as linear ensemble averages over individual structures, i.e. employing Eq. \eqref{eq:ensemble_average}. Therefore, throughout the manuscript the validity of Eq. \eqref{eq:ensemble_average} will be taken as a working assumption until section \ref{sec:time}, where we will discuss the case of time-dependent observables. 

In a reweighting process, the weights are modified until the calculated observables become consistent with corresponding experimental observables, $O^\mathrm{exp}_j$, where $j=1,\ldots,M$. We deonote the reweighted set of weights, $\boldsymbol{w}= w_1,w_2,...,w_N$, the \emph{optimal distribution} \cite{Hummer2015}. Unfortunately, many possible sets of weights can lead to the same averages, and thus to consistency with data. I.e. it is an ill-posed problem, and strategies are therefore needed to regularise the problem, so the most optimal of these distributions can be selected.
%A general solution often assumes that while the reference distribution, $\boldsymbol{w}^0$, is not perfect, it might still be our best initial guess, and can thus be used as starting prior for future refinement. 
There exist several strategies to regularise and solve the resulting optimisation problem \cite{Ravera2016}, and we here focus on three of them: the principle of maximum entropy (MaxEnt, section \ref{sec:maxent}), the principle of maximum parsimony (MaxPars, section \ref{sec:maxpars}) and Bayesian inference (MaxPrior, section \ref{sec:maxprior}). Each method encodes a different philosophy for what it means for a set of weights to be \emph{optimal}, which translates into different optimal distributions after reweighting.

\subsection{Maximum Entropy}\label{sec:maxent}
The principle of maximum entropy (MaxEnt) \cite{Jaynes1957} states that the optimal distribution  is the one adding the least amount of information given some imposed \textit{constraints}, e.g. $\langle O^\mathrm{calc}\rangle = O^\mathrm{exp}$. In other words, the solution with highest entropy and still consistent with the data is considered optimal. In this context, entropy is quantified via the \emph{relative entropy} which we define as the negative of the Kullback-Leibler divergence \cite{Kullback1951}:
\begin{align}\label{eq:entropy1}
S(\boldsymbol{w})= -\sum_{i=1}^N w_i \log\left(\frac{w_i}{w^0_i}\right),
\end{align}
where $w_i^0$ and $w_i$ are the weights of frame $X_i$ before and after reweighting. Other expressions for the entropy are possible \cite{Hansen1991,Skilling1989}, but Eq. \eqref{eq:entropy1} is usually preferred \cite{Rozycki2011,Leung2016,Reichel2018,bottaro2018conformational}. As the distributions of initial and refined weights diverges, $S(\boldsymbol{w})$ becomes smaller, and vice versa. In the simplest case, the set of weights that maximises the entropy is chosen by the only constraint that the weights are normalised as probabilities, i.e. we look for weights for which the function $g_0(\boldsymbol{w}) = 1-\sum_{i} w_i$ is zero. This problem can be solved by constructing a Lagrangian:
\begin{align}\label{eq:lagrangian}
\mathcal{L}(\boldsymbol{w},\lambda_0) = S(\boldsymbol{w}) - \lambda_0 g_0(\boldsymbol{w}),
\end{align}where $\lambda_0$ is a Lagrange multiplier. The solution of the set of equations
\begin{equation}\label{eq:lagrangian_der}
\nabla\mathcal{L}=0
\end{equation}gives the highest entropy weights which fulfil the normalisation condition, which are simply $\boldsymbol{w}=\boldsymbol{w^0}$. It is clear then, that the problem is interesting only when more constraints are introduced. For example, exact consistency with a given set of experimental data can be imposed via functions of the form $g_j(\boldsymbol{w}) = O^\mathrm{exp}_j-\langle O^\mathrm{calc}_j\rangle(\boldsymbol{w})$. The Lagrangian in Eq. \eqref{eq:lagrangian}, in the case of $M$ experimental constraints and the normalisation constraint, becomes:
\begin{align}\label{eq:lagrangian2}
\mathcal{L}(\boldsymbol{w},\boldsymbol{\lambda}) = S(\boldsymbol{w}) - \sum_{j=0}^M\lambda_j g_j(\boldsymbol{w}),
\end{align}where a Lagrange multiplier $\lambda_j$ is introduced for each constraint. The solution of Eq. \eqref{eq:lagrangian_der} for the Lagrangian in Eq. \eqref{eq:lagrangian2} provides weights that ensure consistency between the reweighted trajectory and the experimental data. The optimal weights are uniquely given in terms of the $M$ Lagrange multipliers \cite{Jaynes1978,Roux2013,Boomsma2014,Cesari2018,kofinger2019efficient}:
%, as can be derived by solving the equations $\delta \mathcal{L}/\delta w_i =0$
\begin{align}\label{eq:MaxEnt_2}
w_i(\boldsymbol{\lambda}) =  \frac{w^0_i\exp\left[-\sum_{j=0}^M \lambda_j O^\mathrm{calc}_j(X_i)\right]}{\sum_{k=0}^N w^0_k\exp\left[ -\sum_{j=0}^M\lambda_j O^\mathrm{calc}_j(X_k)\right]},
\end{align}
where $O^\mathrm{calc}_j(X_i)$ is the value of the $j$-th observable, as calculated from the $i$-th frame in the simulation. Note that the denominator just ensures proper normalisation of the weights.

Fitting data tightly to gain full consistency by imposed constraints might, however, be unrealistic and lead to overfitting because of errors in the data and in the model \cite{Roux2013}. Indeed, experimental observables are only known with some limited certainty and, likewise, forward models used to calculate observables from simulations have an associated uncertainty \cite{Cordeiro2017}. Systematic errors in the data lead to further disagreement with the model \cite{Hummer2015,Shevchuk2017,Bonomi2016}, and finally, the finite number of structures $N$ used to compute the theoretical observable (see Eq. \eqref{eq:ensemble_average}) introduces additional error to the estimated averages \cite{Bonomi2016}. All these effects add up to an effective uncertainty for each observable, $\sigma_j$. Therefore it is generally preferable to apply \emph{restraints} rather than constraints, i.e. to impose consistency with the data only within the estimated uncertainty (restraining) rather than exactly (constraining). This translates into relaxing the equality $\langle O^\mathrm{calc}\rangle = O^\mathrm{exp}$ to a similarity $\langle O^\mathrm{calc}\rangle \sim O^\mathrm{exp}$ within a given threshold. 

There are two main strategies to introduce restraints in the MaxEnt method using Lagrange multipliers. The first one consists in modifying the constraints in such a way that the calculated and experimental observables are allowed to differ by some small quantity $\varepsilon$, i.e. $g^\sigma_j(\boldsymbol{w}) = (O^\mathrm{exp}_j + \varepsilon(\lambda_j,\sigma_j))-\langle O^\mathrm{calc}_j\rangle(\boldsymbol{w})$. $\varepsilon$ depends on the effective uncertainty, $\sigma_j$ and the Lagrange multiplier associated to the constraint, $\lambda_j$ \cite{Cesari2016,Cesari2018,Amirkulova2019}. The second way to include experimental errors was introduced by Gull and Daniel \cite{Gull1978} and, in the assumption of normally distributed errors, describes the discrepancy between data and model via the $\chi^2$ distribution: 
\begin{align}\label{eq:chi2}
\chi^2(\boldsymbol{w)} = \sum_{j=1}^M \left( \frac{O_j^\mathrm{exp}-\langle O^\mathrm{calc}_j\rangle (\boldsymbol{w})}{\sigma_j} \right)^2,
\end{align}
The expectation value of $\chi^2$ is equal to the number of degrees of freedom $\nu$, so a reduced $\chi^2$ can be defined such that its expectation value is unity, 
\begin{equation}\label{eq:rchi2}
\chi^2_r =\frac{1}{\nu} \chi^2.
\end{equation}
Experimental restraints can thus be included by imposing the constraint function $g_{\chi^2_r}(\boldsymbol{w}) = \chi^2_r (\boldsymbol{w})-1$ to be zero, with a Lagrange multiplier that we call $\theta ^{-1}$, and whose role we discuss in further detail below. This treatment is correct in the assumption of normally distributed errors and for datasets that guarantee $\chi^2_r=1$. The latter assumption is not safe, as point estimates can deviate from the $\chi^2_r$ expectation value: more details on this are provided in section \ref{sec:challenges}. 
To overcome this problem, the optimal distribution can be obtained by minimising the equation \cite{Hummer2015,Rozycki2011,bottaro2018conformational}
%To overcome this problem, $\theta$ can be treated as an adjustable parameter, controlling how tightly the data should be fitted. Effectively, the optimal distribution is obtained by minimising the equation \cite{Hummer2015,Rozycki2011,bottaro2018conformational}:
\begin{align}\label{eq:MaxEnt_reg}
T_\mathrm{MaxEnt}(\boldsymbol{w}) = \chi^2(\boldsymbol{w}) - \theta S(\boldsymbol{w}),
\end{align}
where $\theta$ is treated as an adjustable parameter rather than a Lagrange multiplier, and controls how tightly the data should be fitted. The solution is then said to be regularised by the entropy, i.e. at a fixed value of $\theta$ the optimal solution is the one with the highest entropy $S(\boldsymbol{w})$ among the ones that minimise the discrepancy with the data, $\chi^2(\boldsymbol{w})$. 

After reweighting, some of the weights may be close to zero and the corresponding structures will thus effectively be ignored in the calculation of the new averages. MaxEnt ensures that the optimised ensemble preserves as many structures from the reference one as possible. In this sense, the entropy term $S$ can be interpreted via a more intuitive quantity, $\phi_\mathrm{eff}$, which represents the effective fraction of frames used in the reweighted ensemble compared to the initial ensemble \cite{bottaro2018conformational}: 
\begin{align}\label{eq:Neff}
    \phi_\mathrm{eff}(\boldsymbol{w})=\exp[S(\boldsymbol{w})].
\end{align}
If the reference distribution is unaltered, $\phi_\mathrm{eff}(\boldsymbol{w}=\boldsymbol{w}_0)=1$. In the opposite extreme all weight is given to a few frames of the simulation and $\phi_\text{eff}(\boldsymbol{w})\ll1$, meaning that the force field is in poor agreement with data and/or sampling is poor. We note that other possibilities exist to estimate the effective size of the sample, e.g. the Kish formula \cite{kish1968survey,rangan2018determination}:
\begin{equation}\label{eq:kish}
\mathcal{K} = \frac{\left(\sum_{i=1}^Nw_i\right)^2}{\sum_{i=1}^Nw_i^2}.
\end{equation}
When a single frame is dominant over all the others, Eq. \eqref{eq:kish} returns $\mathcal{K}=1$, while in a situation where no particular frame is preferred one has $\mathcal{K}= N$. 

\subsection{Maximum Parsimony}\label{sec:maxpars}
The principle of Maximum Parsimony (MaxPars), also referred to as \emph{Occam's razor}, is another strategy that can help choosing among several models consistent with the data. As we shall discuss later, many different interpretations of Occam's razor exist. In this section, the principle is interpreted as follows: the optimal distribution coherent with a given set of data is the one providing the smallest possible ensemble while still fitting the data. The basic idea is that if a model (ensemble) with few parameters (structures) can explain the data, there is no reason to further complicate the model by introducing more parameters. The above principle can for example be quantified by the Akaike information criterion \cite{Akaike1974}:
\begin{align}\label{akaike}
\text{AIC} = 2n - 2\log L,
\end{align}where the optimal solution is found by simultaneously maximising the likelihood, $L$, that quantifies agreement with experimental data, and minimising the number of model parameters, $n$. While the number $n$ is effectively equal to the number of frames in the ensemble, it is customary to associate it to the number of non-zero weights, so $n \ll N$. For normally distributed experimental restraints one has $\log L = -\chi^2(\boldsymbol{w})/2$, so:
\begin{align}\label{eq:MaxPars}
T_\mathrm{MaxPars}(\boldsymbol{w},n) =\text{AIC}= \chi^2(\boldsymbol{w}) + \theta n,
\end{align}
where $\theta=2$ if Eq. \eqref{akaike} is directly applied. $\theta$ was introduced here to emphasise similarity with the MaxEnt methods, i.e. that the difference is in the form of the regularisation term. %As the method minimises the size of the ensemble, it may also be denoted a minimal ensemble method.

The Akaike information criterion was directly applied by Bowerman et al. \cite{Bowerman2017} to find a minimal ensemble of tri-ubiquitin, using MD simulations and SAXS data. Bouma et al. \cite{boura2011solution}, on the other hand, used $\theta$ as a free adjustable parameter to tune the strength of the MaxPars regularisation term. In most methods however, the optimal $n$ is found by increasing it by incremental steps and, for each step, finding the ensemble with lowest $\chi^2(\boldsymbol{w})$. The optimal $n$ is then found when $\chi^2(\boldsymbol{w})$ has converged, either judged by manual assessment \cite{Chen2007,Francis2011} or by an automatic convergence criterion \cite{Cossio2013,Berlin2013,Schneidman-Duhovny2016}, thus avoiding having to set an explicit value for $\theta$. 

\subsection{Bayesian inference or MaxPrior}\label{sec:maxprior}
An alternative method to MaxEnt and MaxPars is provided by Bayesian statistics \cite{Rieping2005,Olsson2013,Hummer2015,kofinger2019efficient}. While MaxEnt regularises the negative log-likelihood by the entropy, and MaxPars by parsimony, Bayesian inference employs the prior for the same goal. For this reason we will refer to this approach as \emph{MaxPrior}. 

In Bayesian inference all available information is expressed by means of probabilities, as captured in Bayes' theorem:
\begin{equation}\label{eq:bayes_theorem}
P(\boldsymbol{w}|\boldsymbol{O^\mathrm{exp}},\boldsymbol{\sigma}) = \frac{P(\boldsymbol{O^\mathrm{exp}}|\boldsymbol{w},\boldsymbol{\sigma})}{P(\boldsymbol{O^\mathrm{exp}},\boldsymbol{\sigma})}P(\boldsymbol{w}).
\end{equation}$P(\boldsymbol{w})$ is called the prior and it quantifies the information known about the system prior to the introduction of experimental data. In case of reweighting, it represents the prior probability associated to the weights, and typically comes from the distribution encoded in an energy function. The term $P(\boldsymbol{O^\mathrm{exp}}|\boldsymbol{w},\boldsymbol{\sigma})$, known as the likelihood $L(\boldsymbol{w})$, represents the probability of measuring the experimental observables, given the set of weights and uncertainties. Assuming normally distributed errors, $L(\boldsymbol{w})$ is given as:
\begin{align}\label{eq:likelihood}
P(\boldsymbol{O^\mathrm{exp}}|\boldsymbol{w},\boldsymbol{\sigma}) =L(\boldsymbol{w})\propto \exp\left(-\frac{1}{2}\chi^2(\boldsymbol{w})\right).
\end{align}Note that this is not the only possible expression for the likelihood. More complicated expressions, even non-analytical ones, can occur if different types of data are combined or when sources of error cannot be assumed to be Gaussian \cite{dutta2018bayesian,pernot2017critical}. The term on the left-hand side of Eq. \eqref{eq:bayes_theorem} is known as the posterior and represents the probability of the weights after experimental data have been considered. Finally, the term in the denominator is treated as a normalisation constant in which case Bayes' theorem takes the form of a proportionality relation:
\begin{align}\label{eq:Bayes_prop}
P(\boldsymbol{w}|\boldsymbol{O^\mathrm{exp}},\boldsymbol{\sigma}) \propto P(\boldsymbol{O^\mathrm{exp}}|\boldsymbol{w},\boldsymbol{\sigma})P(\boldsymbol{w}).
\end{align}Given Eq. \eqref{eq:Bayes_prop}, Bayesian inference defines the optimal distribution by following a two-step procedure: first, the information from data is included, i.e. a likelihood function is defined; second, all other forms of data are included, i.e. a prior on the weights is provided. If we assume the employed force field to represent the best estimate of our prior knowledge on the system, then the prior probability $P(\boldsymbol{w})$ must decrease as $\boldsymbol{w}$ deviates from $\boldsymbol{w^0}$. This observation can e.g. be quantified through the relative entropy $S(\boldsymbol{w})$ from Eq. \ref{eq:entropy1} \cite{Beauchamp2014,bottaro2018conformational,brookes2016experimental}, in which case the prior takes the form:
\begin{align}\label{eq:prior}
P(\boldsymbol{w}) \propto \exp\left(\frac{\theta}{2} S(\boldsymbol{w})\right).
\end{align}The factor $1/2$ is introduced for sake of simplicity of the final result. Other forms of the prior can be used. For example, Gaussian errors are used as priors in several reweighting methods \cite{Fisher2010,Sethi2013,Xiao2014,Ge2018}, while other methods assume a Dirichlet distribution for the weights \cite{Potrzebowski2018}. If the relative entropy term is used as prior, Bayesian inference and MaxEnt approaches result in the same regularised expression. Indeed, by inserting the likelihood (Eq. \eqref{eq:likelihood}) and prior (Eq. \eqref{eq:prior}) into Bayes theorem (Eq. \eqref{eq:Bayes_prop}) and taking the negative logarithm, we obtain a regularised functional, $T_\mathrm{MaxPrior}(\boldsymbol{w})$, that must be minimised to find the optimal distribution, $\boldsymbol{w}$:  
\begin{align}\label{eq:MaxPrior}
-2\log P(\boldsymbol{w}|\boldsymbol{O^\mathrm{exp}},\boldsymbol{\sigma}) \propto T_\mathrm{MaxPrior}(\boldsymbol{w}) =\chi^2(\boldsymbol{w}) - \theta S(\boldsymbol{w}).
\end{align}
%Eq. \eqref{eq:MaxPrior} makes it clear that Bayesian inference is equivalent to an optimisation problem with the negative log-prior as regularisation term, thus the name MaxPrior.  
It is important to note that, differently from the case of MaxEnt where $\theta$ was empirically used to tune the strength of the regularisation term, in a Bayesian framework the interpretation of $\theta$ is clear. Indeed, it accounts fot the uncertainty on the weights, $\sigma_{w^0_i}$, and thus effectively on the force field, i.e. $\theta=(\sigma_{w^0_i})^{-1}$. This means that if the uncertainty on the weights, data and model were known accurately, $\theta$ could be exactly determined. Unfortunately, the uncertainty on the force field parameters, and thus on the weights, is generally unknown and $\theta$ must therefore be treated as a hyperparameter of the model. In practice, $\theta$ plays a double-role: it expresses our trust in the force field and at the same time it can be used to compensate for over- or underestimated experimental errors, as well as errors in the forward model \cite{Hummer2015,Bottaro2018saxs}. See more on the determination of $\theta$ in section \ref{sec:challenges}.

The usefulness of Bayes theorem in refining structural ensembles from simulations and data is clear from the amount of studies on the subject, all including the word \emph{Bayesian} in their title \cite{Fisher2010,Cossio2013,Sethi2013,Beauchamp2014,Molnar2014,Xiao2014,mechelke2014bayesian,Hummer2015,Bonomi2016,Antonov2016,Shevchuk2017,Reichel2018,Potrzebowski2018,Bottaro2018saxs,Ge2018} (see also the overview in \cite{Bonomi2017}). The methodology has become so widespread in the field, that one could argue that researchers should start stating in the title when Bayesian methods are not used, rather than the opposite. We note also that different methods may differ substantially in how and the extent to which they apply the Bayesian formalism including whether priors are defined over all parameters and whether these are integrated out during the procedures.

Several methods combine MaxPars, MaxEnt and MaxPrior approaches (see an overview in Table 1 in the review by Bonomi et al. \cite{Bonomi2017}). Bayesian priors can, e.g., be constructed to prefer minimal ensembles \cite{Fisher2012,Potrzebowski2018}, thus combining MaxPrior with MaxPars. Additional information can also be used together with prior information from the simulation, e.g. secondary structure restraints, like in the case where MD simulated data are fitted into low-resolution cryo-electron microscopy density maps \cite{Kirmizialtin2015}. Therefore, the Bayesian approach should be seen more as a toolbox than an actual principle for selection among ensembles. 

The Bayesian framework also provides an interpretation of the principle of parsimony, which is different from the Akaike Information Criterion, where the model with fewer parameters is considered the simplest one. In a Bayesian setting, the principle of maximum parsimony is a matter of \emph{surprise}: the further a posterior result is from the prior, the more \emph{surprising} it is. This leads to a Bayesian Occam's term: a measure of the distance between the reference distribution (prior) and the optimal distribution (posterior) \cite{MacKay1992, Larsen2018}. The principle is implicitly built into Bayes theorem. In that sense, the $S(\boldsymbol{w})$ term in the Bayesian inference and  MaxEnt methods can be interpreted as an Occam's term, and both these methods could claim to fulfil the principle of maximum parsimony. However, to keep notation clear, we will use the Akaike interpretation of Occam's razor. 

\subsection{Comparing MaxEnt, MaxPars and MaxPrior reweighting}
Different interpretations of the concept of optimal distribution can lead to dramatically different conformational ensembles. Indeed, MaxEnt includes as many configurations of the initial ensemble as possible, while MaxPars tries to include only the ones that are strictly necessary to model the data. The behaviour of MaxPrior, on the other hand, depends on the prior employed. In most approaches however, a term similar to MaxEnt is used, preferring solutions that are consistent with the reference distribution \cite{Bonomi2017}, so that MaxPrior will lead to solutions with as many frames as possible. Therefore, we will only distinguish between MaxEnt and MaxPars in the following, when discussing which method is more appropriate under what circumstances. To our knowledge, no systematic and direct comparison of the different reweighting methods have been made. However, the methods have some clear and important differences that will be discussed below. 

\subsubsection{Interpretation of the results}
A relevant point of comparison between the methods is how easily the results are interpreted. The results of the MaxPars method are easier to interpret and visualise, as the optimal ensemble usually contains only 2-5 representative structures \cite{Pelikan2009,Molnar2014}, whereas the MaxEnt method may result in an optimal ensemble with thousands of structures, whereof many are similar. 

\subsubsection{General applicability}
As argued by Bonomi et al. \cite{Bonomi2017} and Ravera et al. \cite{Ravera2016},  both methods can be used when the system's free energy landscape shows a few distinct and well-defined minima. In that case, the ensemble is represented well by a few structures, one for each free energy minimum and weighted by its corresponding depth (see Fig.~\ref{fig:overview}). On the other hand, if the energy landscape representing the ensemble is flat or high-dimensional, i.e. in case of high-entropy systems, a few structures provide a poor description of the ensemble, even if they may fit the data. So in the case of high-entropy systems, the MaxEnt method should be preferred over MaxPars. A simplistic example of a high entropy system is a regular die\cite{Jaynes1978,Boomsma2014,Ravera2016}. For an unbiased die, all six faces, which represent the states of the system, are equally likely. Let us consider the experimental observation that the average result of throwing the die is 3.5. The application of MaxPars to this problem would lead to the conclusion that the die is well described by two states only, e.g states \emph{3} and \emph{4} with weights $w_3=w_4=0.5$, or states \emph{2} and \emph{4} with weights $w_2=0.25$ and $w_4=0.75$, or any other pair of states consistent with the observation. This result is ambiguous, as many combinations are equally \emph{optimal}, and it is also a poor description of the regular die. MaxEnt, on the other hand, would lead to a uniform distribution for all outcomes. More relevant examples of high-entropy systems are multi-domain proteins with flexible linkers \cite{Vogel2004}, intrinsically disordered proteins \cite{Dyson2005}, and unfolded proteins \cite{lindorff2012structure}, which are all described poorly by a few distinct states. Thus, the MaxEnt method is more versatile, in the sense that it gives reliable results also for high-entropy systems \cite{Ravera2016,Bonomi2017}. The dimensionality of the system also plays an important role. For reasonably small systems, a description in terms of low-dimensional free-energy landscapes might be possible and relevant. I.e. a few structures can adequately represent the system. However, for high-dimensional biological systems the assumption that a few coordinates, or states, can properly describe the collective dynamics of the molecule might be ventured. Recent studies have highlighted how also the dynamics of some fast-folding proteins might be too complex to be represented in fewer than $\sim 10$ dimensions \cite{rodriguez2018computing}. Therefore, we suggest that MaxEnt, or related methods, should be considered the method of choice when no safe assumptions can be made about the free energy landscape of the ensemble. 

\subsubsection{Imperfect force fields}
\begin{figure}[tbp!]
\begin{center}
\includegraphics[width=1.0\linewidth]{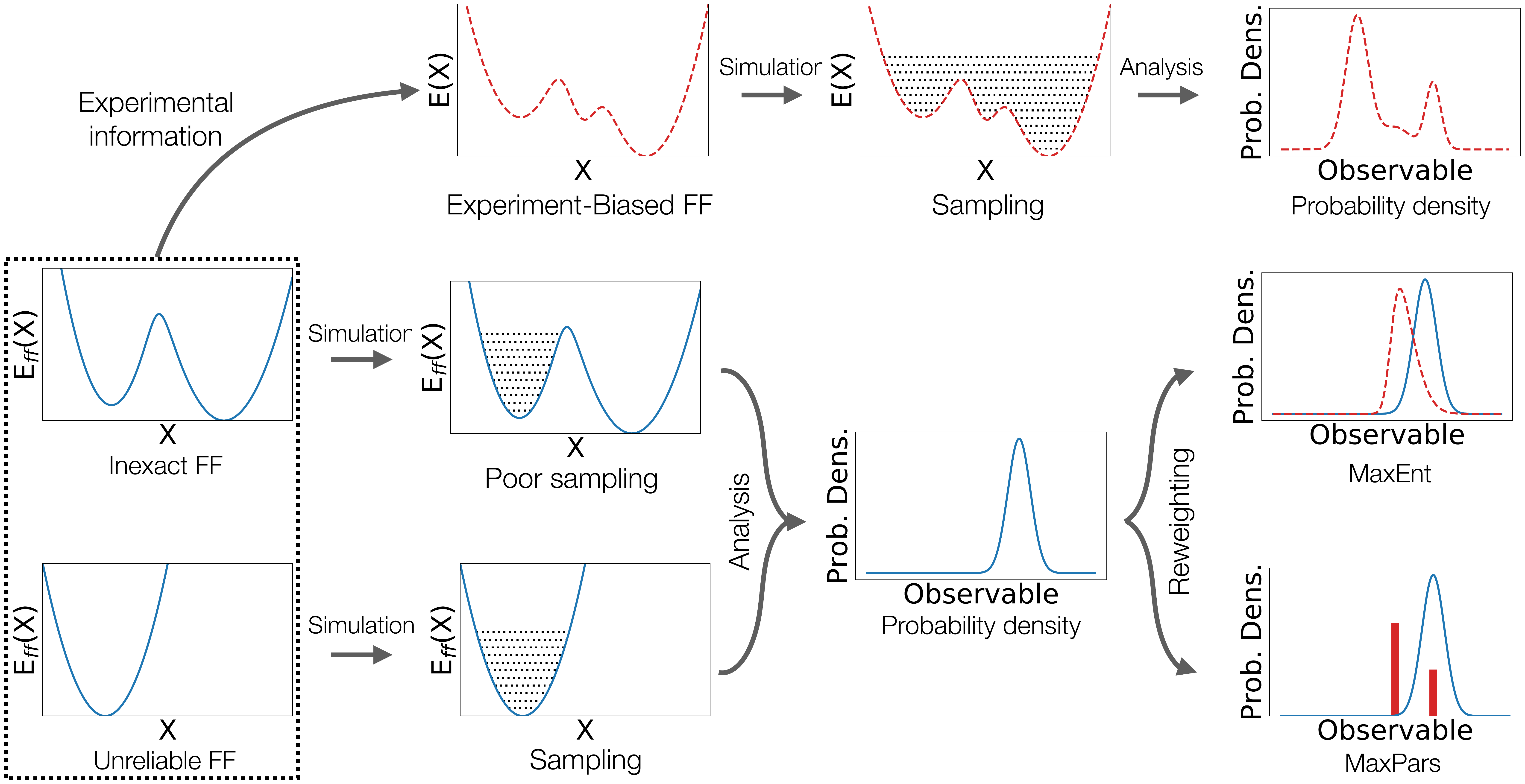}
\end{center}
\caption{Lower two rows: The two cases of poor sampling of an inexact force field and the one of full sampling of an unreliable force field can both lead to a result where the distribution of an observable of interest is poorly estimated. The performance of subsequent MaxEnt/MaxPrior and MaxPars reweighting differ, as a poor prior inevitably leads to a poor reweighted distribution in the case of MaxEnt/MaxPrior, while MaxPars can still select some relevant conformations from the initial unreliable pool. Top row: In both cases, experiment-biased simulations can help alleviate the problem.}
\label{fig:poorFF}
\end{figure}

The choice of the reweighting method also depends on the quality of the simulations used to generate the reference distribution $\boldsymbol{w^0}$. Indeed, the results of MaxEnt and MaxPrior (with e.g. Gaussian \cite{Fisher2010,Sethi2013,Xiao2014,Ge2018} and Dirichlet priors \cite{Potrzebowski2018}) reweighting will strongly depend on the quality of the simulation, because the optimal distribution is expected not to deviate too much from $\boldsymbol{w}^0$. On the other hand, MaxPars reweighting depends less on the quality of the initial distribution, as it freely picks the simulated structures that best represent the data. This decoupling from the initial simulation is both the strength and the weakness of the MaxPars method. It implies that, in principle, a realistic reference distribution is not necessarily preferred over a poor one, as long as some good representative structures have been sampled (Fig. \ref{fig:poorFF}). Consequently, approximate but efficient simulation methods, e.g. Monte Carlo sampling with implicit water, can be used in conjunction with the MaxPars method (e.g. Rosetta \cite{Alford2017}), allowing for a fast exploration of the conformational space and better overall sampling without the need for enhanced sampling methods. Such approximate simulation methods are evidently less suitable for the MaxEnt and MaxPrior reweighting and therefore, more costly explicit solvent MD simulations should be employed. On the other hand, MaxPars does not benefit much from the qualified prior information of a good force field. We want to stress here that one of the main advantages of using accurate simulations as a prior is when the provided experimental dataset is sparse. In that case, the refined ensembles obtained from MaxEnt/MaxPrior are generally more reliable than MaxPars, as they are regularised by the simulated model.

\subsection{Numerical challenges}
Reweighting methods, as we presented them here, are essentially optimisation problems. As such, they are cursed by the problem of dimensionality \cite{bellman2015adaptive} or, in other words, the higher the dimension of the parameter space the more computationally challenging it becomes to find the true minimum of the functional of interest. Several strategies have been employed to reduce the computational burden of reweighting. 

In MaxEnt, for example, the principal numerical challenge is to find the set of weights that minimise Eq. \eqref{eq:lagrangian2}. This can become highly nontrivial when $N$ is large, e.g. for tens of thousands of structures. The computational complexity can nonetheless be alleviated by optimising the values of the $M$ Lagrange multipliers (Eq. \eqref{eq:MaxEnt_2}) rather than the $N$ weights, as usually $M \ll N$ \cite{Rozycki2011,Hummer2015,bottaro2018conformational}. We note however that, depending on the implementation, the optimisation of the $N$ weights can be faster than the one of the Lagrange multipliers \cite{kofinger2019efficient}. Moreover, the functional in Eq. \ref{eq:MaxPrior} is strictly convex for finite $\theta$ \cite{agmon1979algorithm}, so the corresponding optimisation can be carried out in a straightforward fashion by means of highly efficient routines such as  limited memory Broyden-Fletcher-Goldfarb-Shanno (L-BFGS) algorithm \cite{malouf2002comparison}. 
%Typically, highly efficient routines, e.g. limited memory Broyden-Fletcher-Goldfarb-Shanno (L-BFGS) algorithm \cite{malouf2002comparison} or simulated annealing \cite{kirkpatrick1983optimization}, are implemented to solve the optimisation problem. We note that, in the case of MaxEnt with finite $\theta$, the function to be minimised is strictly convex \cite{agmon1979algorithm}, so its optimisation with methods such as L-BFGS should result straightforward. 

In MaxPars, the numerically challenging part is provided by the test of the $N!/(n!(N-n)!)$ combinations of $n$-sized ensembles out of $N$ structures. A naive approach where all the possible combinations are tested is intractable for realistic situations as, for example, in the case of $n=5$ and $N=1000$ the number of combinations becomes $C(n,N) \sim 10^{13}$. Therefore, the best fitting combinations of $n$ structures has to be determined with some other strategies, e.g. via Monte Carlo minimisation \cite{Bernado2007} or more complex algorithms such as the one implemented in ASTEROID \cite{Nodet2009}. Other strategies limit the search to the subset of structures that best match the experimental data individually \cite{Schneidman-Duhovny2016}, which however violates the principle that the goodness of fit should only be assessed only against the ensemble average (Eq. \eqref{eq:ensemble_average}), and not against the fit to single structures. 

As a final remark, we note that both MaxPars and MaxEnt methods can benefit from clustering, where structures are grouped according to their structural resemblance, to reduce $N$ before reweighting \cite{Rozycki2011,Sethi2013,Xiao2014}.

\section{Experiment-biased simulations} \label{sec:on-the-fly}
In section \ref{sec:reweighting} we focused on three possible principles, MaxEnt, MaxPars and MaxPrior, to reweight the results of simulations \emph{a posteriori}. However, as we argued in section \ref{sec:introduction}, this is not the only possible approach. Instead of changing the simulated ensemble after the simulation has been performed, the simulation can be guided by adding a bias to the underlying force field, $E_\mathrm{ff}(X)$, employed in the simulation to ensure consistency with the experimental data. In this section we provide an overview of the methods that have been applied with this goal, which are either based on the MaxEnt principle (section \ref{sec:exp_bias:maxent}), on the addition of empirical energy terms (section \ref{sec:exp_bias:empirical}), or on the MaxPrior principle (section \ref{sec:exp_bias:maxprior}). Finally, in section \ref{sec:compare} we compare and discuss the differences between experiment-biased simulations and \emph{a posteriori} reweighting strategies, discussing their possible advantages and shortcomings. 

\subsection{Maximum Entropy}\label{sec:exp_bias:maxent}
In section \ref{sec:maxent}, we discussed how Maximum Entropy can be used to obtain consistency between data and simulations by optimising a set of Lagrange multipliers (see Eq. \eqref{eq:MaxEnt_2} and Ref. \cite{Cesari2018}). The mathematical formulation of the principle makes it immediate to extend it to the problem of experiment-biased simulations. Starting from an unbiased force field $E_\text{ff}(X)$ (Fig. \ref{fig:overview}), the idea is to perturb it as little as possible under the constraint/restraint of consistency with some given experimental data. Maximum Entropy principle can be expressed in terms of the probability of a conformation $X$ given the new force field $P(X)\propto\exp(-\beta E(X))$, where $\beta=(k_\text{B}T)^{-1}$ with $k_\text{B}$ being the Boltzmann constant and $T$ the temperature of the system, and the probability of the same configuration obtained from the unbiased force field $P_\mathrm{ff}(X)\propto\exp(-\beta E_\text{ff}(X))$ \cite{Roux2013,Amirkulova2019}:
\begin{align}
S[P(X)] = - \int dX P(X) \log\left(\frac{P(X)}{P_\mathrm{ff}(X)}\right).
\end{align}
Practical examples of this approach are provided by experiment directed simulations (EDS) \cite{White2014} and experiment directed metadynamics (EDM) \cite{White2015}. The effective error from experiment and forward model can also be accounted for in the Lagrangian framework, thus only restraining the solution within the given uncertainty \cite{Amirkulova2019,Cesari2016}. 

\subsection{Empirical energy terms}\label{sec:exp_bias:empirical}
Rather than approaching with MaxEnt, experimental restraints can be directly included as empirical penalty terms \cite{Lindorff-Larsen2005}:
\begin{align}
E(X) &= E_\mathrm{ff}(X) + E_\mathrm{exp}(\boldsymbol{\langle O^\mathrm{calc}\rangle},\boldsymbol{O^\mathrm{exp}},\boldsymbol{\sigma}) \nonumber \\
 &= E_\mathrm{ff}(X) + \sum_{j=1}^M \theta_j h_j(\langle O^\mathrm{calc}_j \rangle,O^\mathrm{exp}_j,\sigma_j),
\end{align}
where the sum is carried out over $M$ restraints. $\theta_j$ are the force constants associated to each restraint, while the specific choice for the functional form of each $h_j$ depends on the distribution of the corresponding observable. The constraints are usually expressed as squared residuals \cite{Lindorff-Larsen2005,Bonomi2016} or, when errors are taken into account, as $\chi^2$ terms \cite{Bonomi2017} (Eq. \eqref{eq:chi2}).

For concreteness and simplicity, we assume here normally distributed experimental observables. We also assume that the same force constant can be used for all observables. This is a good assumption, e.g. when the observables come from SAXS data \cite{Shevchuk2017}, but cannot generally be assumed when more than one experimental techniques are combined. With this assumption, the energy function reduces to: 
\begin{align}\label{eq:on-the-fly}
E(X) = E_\mathrm{ff}(X) + \theta\chi^2(\boldsymbol{\langle O^\mathrm{calc}\rangle},\boldsymbol{ O^\mathrm{exp}},\boldsymbol{\sigma}). 
\end{align}
Just as in reweighting approaches, parameter $\theta$ in Eq. \eqref{eq:on-the-fly} takes into account unknown uncertainties of the force field parameters as well as unknown or imperfectly determined errors in the data and the forward model. Therefore, $\theta$ is generally not known and determining it is one of the key challenges of the method \cite{Hummer2015,kofinger2019efficient}.

Another issue in the practical implementation of this approach is that the average $\boldsymbol{\langle O^\mathrm{calc}\rangle}$ can only be determined \emph{after} the simulation is over. Thus the restraints need to be applied iteratively \cite{Boomsma2014}. As an alternative approach to obtain an ensemble averages at each point in the simulation, replica methods have been introduced \cite{Vendruscolo2007,Burnley2012,Levin2007,Hummer2015}, where $N$ independent replicas of the same system are simulated and observables are calculated as ensemble averages from these. Interestingly, replica experiment-biased methods mathematically converge to a Maximum Entropy \emph{constrained} solution as $N\rightarrow\infty$ and $\theta\rightarrow\infty$, as discussed by Pitera \& Chodera \cite{Pitera2012}, Roux \& Weare \cite{Roux2013} and Cavalli et al. \cite{Cavalli2013}, and reviewed by Boomsma et al.~\cite{Boomsma2014}. Very recently,  K{\"o}finger et al.~\cite{kofinger2019efficient} combined reweighting and experiment-biased methods to simultaneously ensure a large ensemble (as in MaxEnt) is considered and that all relevant states are visited by adding a biasing energy term to the force field.

\subsection{Bayesian inference}\label{sec:exp_bias:maxprior}
Just as in the case of MaxEnt, Bayesian inference, or MaxPrior principle (see section \ref{sec:maxprior}), can be also employed to bias \emph{a priori} molecular simulations rather than just reweighting them \emph{a posteriori}. The method known as Metainference \cite{Bonomi2016} is an implementation of this principle: it employs a Bayesian approach to quantify how much the prior is modified by the introduction of noisy and heterogeneous sources of data. In its essence, the strategy works by running $N$ replicas of the system and guiding the sampling by means of a log-posterior scoring function:
\begin{equation}\label{eq:metainf}
s(X,\boldsymbol{\sigma}) = - \sum_{i=1}^N \log P(X_i,\sigma_i) + \Delta^2(X) \sum_{i=1}^N \frac{1}{2\sigma^2_i},
\end{equation}where $\boldsymbol{\sigma}$ takes into account all the sources of errors, $P(X_i,\sigma_i)$ is the prior and the factor $\Delta^2(X)$ estimates the deviation between the predicted observables and the experimental ones. It can be shown that in the single-replica limit, $N=1$, Eq. \eqref{eq:metainf} reduces to Eq. \eqref{eq:MaxPrior}. Therefore, the log-prior plays the role of an effective entropy term, while the second component of $s(X,\boldsymbol{\sigma})$ is a $\chi^2$ term computed over the replicas. It is interesting to notice here that the equivalence between Eq. \eqref{eq:metainf} and Eq. \eqref{eq:MaxPrior} is valid only in the $\theta=1$ case: this comes from the fact that the method proposes a model to take into account all the sources of error \cite{Bonomi2016}, which means that $\theta$ is not expected to be a free parameter anymore. We notice that Eq. \eqref{eq:metainf} is only valid for Gaussian error sources, and more complicated and complete expressions can be obtained in the general case (see the discussion in section Materials and Methods in Ref. \cite{Bonomi2016} and Ref. \cite{lohr2019practical}). 

Metainference has been effectively combined with Metadynamics \cite{laio2008metadynamics,barducci2011metadynamics,sutto2012new} in its parallel bias formulation \cite{pfaendtner2015efficient,lohr2019practical}: this synergy enables one to explore the configuration space in an efficient way while simultaneously sampling conformations that are coherent with experimental data. %More on this in section \ref{sec:adapt}.   

\subsection{Comparing reweighting with experiment-biased methods}
\label{sec:compare}

\subsubsection{Adaptability}\label{sec:adapt}
Reweighting methods can be used with many different types of simulations and force fields, as the reweighting process is independent from the simulation and sampling (Fig. \ref{fig:overview}). This makes it a rather adaptable module-like tool, with the input being the trajectory and the experimental data only \cite{bottaro2018conformational,Cesari2018,rangan2018determination}. Also, new experimental data can easily be incorporated in the reweighting process without having to re-run the conformational sampling. In the experiment-biased simulations, the implementation is more specific, as the empirical energy term is an integrated part of the simulation \cite{Shevchuk2017}. Decisions about types of experiments and force field constants have therefore to be taken before the simulation. It is still possible, however, to reweight experiment-biased simulations \emph{a posteriori} to remove or add sets of experimental data, though the procedure is technically somewhat more challenging than reweighting MD trajectories; indeed, the applied experimental bias has to be estimated and subtracted from the simulation before further reweighting \cite{Hummer2015,rangan2018determination}. For the same reason, experiment-biased simulations are often not carried out together with enhanced sampling techniques. A notable exception is provided by metadynamics \cite{bonomi2016metadynamic}, which has been combined with metainference \cite{Bonomi2016,lohr2019practical} to increase its efficiency, and in principle metadynamics (and other enhanced sampling methods) could be used in other methods for experiment-biased simulations.

\subsubsection{Forward models}
When a trajectory is reweighted \emph{a posteriori}, the forward model is only evaluated on the frames that are to be reweighted, which are typically only a small fraction of the frames generated during the simulation. Also, as long as the observables are calculated according to Eq.~\eqref{eq:ensemble_average}, the calculations are done independently of one another and may thus be easily parallelised. For these reasons, the forward model can be of high complexity (e.g. quantum calculations can be carried out on the ensemble structures \cite{ianeselli2018atomic}). This is not the case, however, of experiment-biased simulations: when implemented within a molecular simulation, forward models are evaluated and differentiated at (almost) every step, so they have to be sufficiently simple to assure computational efficiency \cite{rangan2018determination} and their gradients have to be known analytically. Therefore, for complex forward models reweighting might be more applicable than \emph{a priori} experimental bias. In some cases, such as for NMR chemical shifts, it is possible to employ a fast forward model \cite{kohlhoff2009fast} that is almost as accurate as more refined and complex models \cite{shen2010sparta+}, which would be too complex to be computed at each step in a simulation. In other cases, it might not be possible to derive sufficiently computationally efficient and accurate forward models. We suggest that a possibility would be to use simpler and less accurate models to bias the simulations and then reweight \emph{a posteriori} the simulation with the more realistic models.

\subsubsection{Imperfect force fields}
Reweighting methods and experiment-biased simulation methods may in practice perform differently in cases where the force field provides a relatively poor description of the system's conformational space. With a poor force field, some relevant states may be rarely or never visited when running an unbiased simulation (Fig. \ref{fig:poorFF}). Consequently, reweighting methods may fail in predicting the correct average of an observable. An indication that this is happening is usually provided by the fraction of effective frames $\phi_\text{eff}$ becoming close to 0 and by the reweighted distributions of relevant observables being skewed towards the experimental average (Fig. \ref{fig:poorFF}). In principle, this can be overcome by sufficient sampling, but it might be computationally very expensive and practically unrealistic. Experiment-biased force fields, on the other hand, can better provide reasonable results even when a rather poor unperturbed force field is used as basis, as the empirical energy term will alter the energy landscape such that even relevant states with high energies in the unbiased force field become reachable (Fig. \ref{fig:poorFF}). The effect of the added empirical energy term can be monitored by comparison with a control simulation without the additional energy term ($\theta=0$ in Eq. \eqref{eq:on-the-fly}).

\section{Force field optimisation} \label{sec:FF}
In sections \ref{sec:reweighting} and \ref{sec:on-the-fly} we described techniques to refine simulations in a system-specific manner by including experimental data. From a different perspective, substantial progress has been made when using experimental data to improve the force field as a general and transferable predictive tool. In this section we focus on this point and review some of the fundamental advancements in the subject. 

\subsection{Background on force field parametrisation} 
As widely discussed in literature and assessed by empirical knowledge, the quality of the physical description provided by force fields is fundamental for the accuracy of biophysical simulations. The reliability of these simulations indeed depends critically on the ability of the underlying physical description to effectively model all the relevant inter-atomic interactions. After decades of force fields development \cite{jorgensen1996development,oostenbrink2004biomolecular,mackerell1998all,cornell1995second}, MD simulations have reached a high level of reliability and the ever-growing amount of experimental observations calls for a systematic and detailed comparison of theoretical predictions, coming from MD simulations, with the available data.

Unfortunately, force fields do not always provide results that are in perfect agreement with experimental findings. To understand the reasons and the sources of these emerging discrepancies, we shall first recall how a force field is usually designed. For a comprehensive introduction on the subject we refer the reader to Chapter~1 of this book, and here we instead focus on how experimental data may be used in force field parameterization.

The typical force field is composed by two fundamental elements: (i) A functional form $E(X)$. This element embeds our physical understanding of molecular processes by providing a classical parametrisation of the Born-Oppenheimer energy surface \cite{tuckerman2010statistical}. The typical functional form of the force fields used for biomolecular simulations (with minor re-adjustments among the different interpretations) is given by \cite{Guvench2008}:
% \begin{align}\label{eq:force_field}
% E(X) &= E_{\text{bonded}}(X)+E_{\text{non-bonded}}(X) \nonumber \\
% &= \sum_{\text{bonds}}k_b(r-r_0)^2 + \sum_{\text{angles}}k_{\theta}(\theta-\theta_0)^2 + \sum_{\text{dihedrals}}k_{\chi}[1+\cos(n\chi - \chi_0)]  \nonumber \\
% &+\sum_{\text{improper}}k_\phi [1+\cos(2\phi)] + \sum_{i,j\in\text{non bonded}} \left[  \frac{A_{ij}}{r_{ij}^{12}} - \frac{B_{ij}}{r_{ij}^6} + \frac{q_iq_j}{r_{ij}}\right] ,
% \end{align}
\begin{align}\label{eq:force_field}
E(X) &= E_{\text{bonded}}(X)+E_{\text{non-bonded}}(X) \nonumber \\
&= \sum_{\text{bonds}}k_b(r-r_0)^2 + \sum_{\text{angles}}k_{\theta}(\theta-\theta_0)^2 + \sum_{\text{dihedrals}}k_{\phi}[1+\cos(n\phi - \phi_0)]  \nonumber \\
&+ \sum_{i,j\in\text{non-bonded}} \left[  \frac{A_{ij}}{r_{ij}^{12}} - \frac{B_{ij}}{r_{ij}^6} + \frac{q_iq_j}{r_{ij}}\right] ,
\end{align}
where $X$ is a configuration of the system and $r$, $\theta$ and $\phi$ are functions of the atomic coordinates and $q$ are atomic charges. $k_b$, $k_\theta$ and $k_\phi$ denote the different strengths of the interactions and are usually tensors, as they depend on the specific group of atoms involved in the interaction; (ii) A set of parameters $\boldsymbol{\xi} = (r_0, \theta_0, \phi_0, k_b, k_\theta, k_\phi, \ldots)$. The functional expression of the force field depends on the choice of these parameters which set, for example, the strength of interactions, equilibrium distances and angles. To determine their values, it is necessary to fit them against known experimental and quantum mechanical (i.e. ab initio) properties. The specific choice of these properties depends on the philosophy underlying the force field development.

Historically, force field development has always been a daunting task because of its technical complexity and the amount of time, experimental data and simulations required \cite{ponder2003force, zhu2012recent,dauber2019biomolecular,hagler2019force}. To give an example of this, let us focus on the AMBER class of force fields. In its first version \cite{cornell1995second}, bonded angles were fit to the vibrational frequencies of single amino acids or small molecules, in order to reproduce experimental frequencies; fixed charges were fit to reproduce the results of quantum calculations \cite{bayly1993well,jakalian2000fast,jakalian2002fast}. Lennard-Jones parameters, instead, were set in such a way to reproduces enthalpies of vaporisation and densities in organic liquids  \cite{jorgensen1996development}. Finally, dihedral and torsional angles were fit to reproduce quantum calculations of single amino acids or experimental barrier heights of small molecules. With the increasing quality of experimental and ab initio data, modern and widely used versions of the AMBER force fields \cite{lindorff2010improved} have reached a high level of complexity. Nonetheless, the underlying general philosophy remained the same: fit the force field parameters to ab initio and experimental data of single amino acids or small molecules and compare the results against data available for larger systems. Force field ff19SB, obtained by only employing ab initio simulations, constitutes a notable exception \cite{tian2019ff19sb}. Despite its historical significance and great success, this procedure shows some important limitations: (i) It is difficult, within this approach, to improve the force field parameters by making use of discrepancies with respect to experimental data on large biomolecules; (ii) errors induced by the forward models used to calculate experimental observables are not consistently taken into account, and errors on the force field parameters are not estimated and/or not used. In the last decade, many new strategies for force field parametrisation have been developed, that less strictly follow the traditional approaches and instead embrace a more Bayesian approach  \cite{norgaard2008experimental,best2009optimized,cailliez2011statistical,rizzi2012uncertainty,angelikopoulos2012bayesian,wang2012systematic,wang2014building,wu2016hierarchical,wang2017building}. These efforts, combined with the growing interest in automated force field parametrisation (as exemplified by the development of the ForceBalance framework \cite{wang2012systematic,wang2014building,wang2017building}),  partially solve the issues reported in point (ii). A deeper discussion on point (ii) will be left for section \ref{sec:challenges}. In this section we focus on point (i) or, more precisely, describe and summarise the strategies available in the literature to optimise force field parameters using discrepancies between MD simulations and experimental observations on longer peptides or even entire proteins. As we will see, these methods are tightly connected to common reweighting algorithms (see section \ref{sec:reweighting}) but, when applied to a wide set of molecules, provide a an answer of more general purpose than the sole ensemble refinement. 

\subsection{Refining Protein and RNA force fields}\label{sec:ff_opt}
NMR data have been widely and successfully used in combination with ensemble reweighting strategies of proteins \cite{Leung2016,bottaro2018conformational,Beauchamp2014,brookes2016experimental,Xiao2014,Ge2018,Bottaro2018saxs,Chen2007,bonvin1994time,huang2009ensemble,lange2008recognition,olsson2014probabilistic,esteban2010refinement,scheek1991structure,mantsyzov2014maximum,mantsyzov2015mera,Nodet2009,graf2007structure,cesari2019fitting,vasile2019determination}; other approaches employed various sources of experimental data, e.g. for small molecules. This fact opens to the appealing possibility to use NMR data on full proteins to directly optimise force fields. Such an approach is more general than ensemble reweighting, in the sense that experimental NMR data can be collected for a larger number of proteins and used to increase the quality of the force field rather than optimising the description of a single system of interest. To understand how this could be possible in practice, we need first to realise that the general scheme of ensemble reweighting methods basically comprises of three steps: (i) Generation of an ensemble of structures with a given force field; (ii) Calculation of experimental observables from the ensemble; (iii) Determination of weights, associated to each structure, that better describe the experimental observables.

\begin{figure}[p!]
\begin{center}
\includegraphics[width=\textwidth]{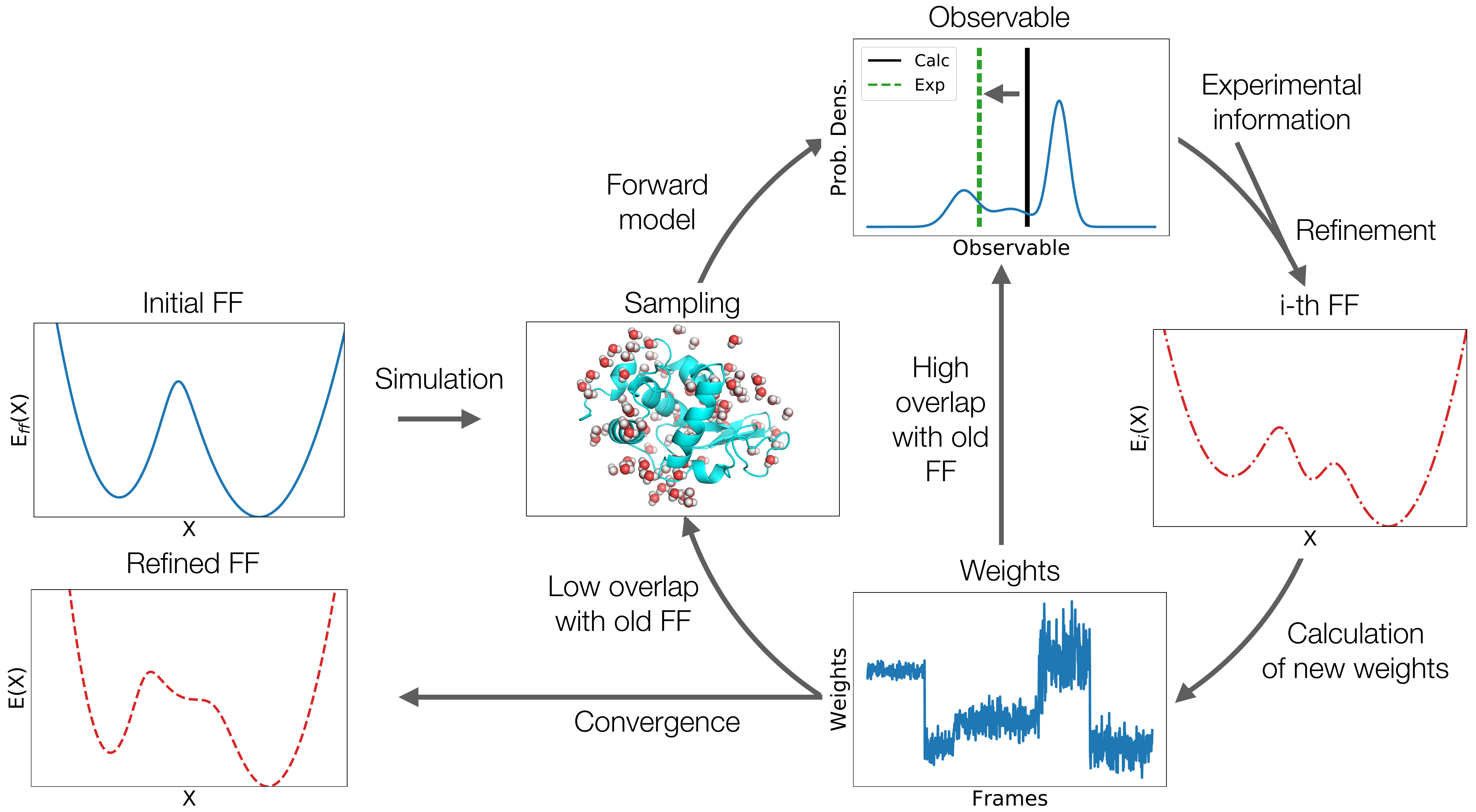}
\end{center}
\caption{Schematic representation of the force field optimisation process. First, an initial force field $E_\text{ff}(X)$ is employed to sample an ensemble of configurations of some systems of interest. A forward model is then applied to the computed frames, to reproduce the known experimental average of a set of relevant observables. If the computed and the experimental observables disagree, further experimental information is used to refine the current force field parameters and defining a new force field $E_i(X)$. A weight is associated to each frame obtained from the previous sampling using Eq.  \eqref{eq:reweight_ff}. If the overlap between the old and the new force field is high (i.e. most of the weights do not change substantially) the new weights are used to refine the estimation of the observable's average, which in turn is used to further refine the force field parameters and to estimate new weights. If the overlap is instead small, a further round of conformational sampling is carried out with the new force field $E_i(X)$. The process is repeated until some convergence criterion is satisfied.}
\label{fig:ff_optimization}
\end{figure}

If this procedure is repeated for a large set of proteins, the information carried by the optimised weights can be in principle used to increase the quality of the underlying physical model, i.e. the values of the force field fit parameters, rather than just the single molecule ensemble (Fig. \ref{fig:ff_optimization}). We note, however, that this approach is less flexible than, e.g., MaxEnt where the weights of all the simulation frames can be fine tuned. When optimising a force field, one is inevitably constrained by its functional form and only limited adjustments can be made. 

In the next sections we review the different strategies that have been developed to compute and use the weights in force field refinement for proteins (section \ref{sec:opt:proteins}) and RNA (section \ref{sec:opt:rna}).

\subsubsection{Proteins}\label{sec:opt:proteins}
Let us assume we can access a given set of $N$ snapshots $X_i$ obtained from a sampling strategy of choice (MD or Monte Carlo) and generated using a given force field $E_\text{ff}$, defined by a set of parameters $\boldsymbol{\xi}_0$. These snapshots can be used to calculate the ensemble average of some observables of interest, $\langle O^\mathrm{calc}_j \rangle$, for which we possess experimental information $O^\mathrm{exp}_j$. The goal is to determine a set of force field parameters $\boldsymbol{\xi}$ that decreases the discrepancy between experimental and simulated averages. This discrepancy can be estimated via the $\chi^2$ (Eq. \eqref{eq:chi2}), whose minimisation with respect to the force field parameters would maximise the compatibility between simulations and the experiments of choice. The search for the $\chi^2$ minimum can in principle be done in a brute force fashion, by infinitesimally perturbing the set of force field parameters by a quantity $\delta\xi_i$ and re-simulating the system of interest. However, this strategy is particularly inefficient, as it corresponds to a search in a high-dimensional parameter space, where the evaluation of each new trial force field requires a full re-sampling of the system's conformational ensemble. To avoid this, it is possible to use some ideas coming from statistical mechanics. In principle, if we knew the \emph{exact} force field $E_\text{exact}$ describing our system, we could reweight each configuration using the Boltzmann relationship:
\begin{equation}\label{eq:bayes_MD}
w_i^\text{exact} = w^0_i e^{-\beta(E_\text{exact}(X_i) - E_\text{ff}(X_i))}.
\end{equation}
Eq. \eqref{eq:bayes_MD} is the same idea behind the method of free energy perturbation \cite{zwanzig1954high} devised by Zwanzing to determine the changes in the free energy of a system when a perturbing potential is introduced. In this case, force field $E_\text{ff}$ plays the role of the unperturbed potential of the system, while the exact force field $E_\text{exact}$ is the contribution to the total energy provided by the perturbation. Even though $E_\text{exact}$ can never be known, Eq. \eqref{eq:bayes_MD} can be used as the basis of an efficient optimisation strategy \cite{norgaard2008experimental,li2010nmr,li2011iterative,chen2018learning}. In this method, a force field $E_\text{old}$ is iteratively optimised to obtain a new one, $E_\text{new}$, that is used to recompute the weights of the parent MD simulation via:
\begin{equation}\label{eq:reweight_ff}
w_i^\text{new} = w_i^\text{old} e^{-\beta(E_\text{new}(X_i) - E_\text{old}(X_i))}.
\end{equation}
The new weights $w_i^\text{new}$ are used to refine the estimation of the simulated average by means of Eq. \eqref{eq:ensemble_average}. $E_\text{new}$ force field is obtained by minimising the $\chi^2$, where the specific strategy for minimisation can change (e.g. Levenberg-Marquardt procedure \cite{norgaard2008experimental,teukolsky1992numerical}, simplex minimisation \cite{li2010nmr} or simulated annealing followed by simplex minimisation  \cite{li2011iterative}).  The strength of the reweighting strategy in Eq. \eqref{eq:reweight_ff} is that it does not require one to sample new conformations every time the force field parameters are updated. Nonetheless, it only works under the assumption that minimal perturbations of the force field are introduced by the $\chi^2$ minimisation and therefore that the snapshots $X_i$ could have been reasonably sampled also by means of the $E_\text{new}$ force field. For this reason, a sensible overlap between $E_\text{old}$ and $E_\text{new}$ force fields is expected to exist, and particular care is dedicated to the assessment of this overlap \cite{bruschweiler1995collective, li2011iterative}, which can be quantified, e.g. through  $\phi_\text{eff}$ (Eq. \eqref{eq:Neff}). A re-sampling of the full conformational ensemble needs to be repeated every time the overlap between the two force fields becomes too small: when $\phi_\text{eff} < \varepsilon$, with $\varepsilon\ll 1$ being a given threshold, one can argue that the overlap between force fields is too small to allow a further round of optimisation. As described, this strategy is based on a method designed to compute free energy differences between different energy function. Thus, for future applications other and more general methods could be employed including Bennett Acceptance Ratio \cite{bennett1976efficient} and its multi-state extension \cite{shirts2008statistically}. These methods may likely also be fruitfully combined with the use of (adaptive) surrogate models \cite{angelikopoulos2012bayesian}, though work remains to be done to make such models accurate and efficient for high dimensional problems.

Norgaard et al. \cite{norgaard2008experimental} introduced the method discussed above to optimise a force field for coarse-grained simulations of unfolded proteins. Reference data were collected from paramagnetic relaxation enhancement NMR, useful to determine long-range effects in unfolded proteins \cite{gillespie1997characterization1,lietzow2002mapping,yi2000nmr,teilum2002transient}. In the first round of optimisation, all the interaction parameters of the coarse-grained force field were set to zero for sake of simplicity, while conformational ensembles are generated using Metropolis Monte Carlo. The method was applied to the $\Delta 131\Delta$ fragment of staphylococcal nuclease \cite{gillespie1997characterization1,gillespie1997characterization2} to obtain reproducible and consistent results. 

Li and Br\"{u}schweiler \cite{li2010nmr} applied the strategy from Norgaard et al.~\cite{norgaard2008experimental} to all atom simulations to derive the Amber ff99SBnmr1 force field starting from backbone dihedral angle potential of Amber ff99SB \cite{hornak2006comparison}. This was done by employing as observables the time-averaged chemical shifts of $\textrm{C}_\alpha$, $\textrm{C}_\beta$ and $\textrm{C}'$ carbon atoms of 4 trial proteins and subsequently benchmarking it against a test set of 18 proteins of different topologies. The obtained results show an average improvement of the comparison with the experimental data for the proteins in the test set, and was later refined further \cite{li2011iterative}.

More recently Chen et al. \cite{chen2018learning} extended the method from Norgaard et al.~\cite{norgaard2008experimental} to use Markov State Models (refer to section \ref{sec:time:maxlike} for more details on Markov State Models) as an intermediate step to calculate observables from simulations and parameterize the conformational landscape. Despite the different framework, the general philosophy of the method, named ODEM (Observable-driven Design of Effective Molecular models), remains unaltered. ODEM has been applied to design a $\textrm{C}_\alpha-\textrm{C}_\beta$ coarse-grained model of protein FIP35 which is able to reproduce relevant pair-distance distributions measured by FRET. 

\subsubsection{RNA}\label{sec:opt:rna}
Molecular simulations may also be used to study the conformational landscapes and dynamics of RNA molecules. Unfortunately, the precision of RNA force fields is still limited and not comparable to the one reached by force fields designed for proteins \cite{condon2015stacking,bergonzo2015highly,kuhrova2016computer,bottaro2016free}. The approach of integrative structural biology therefore becomes a powerful tool to enhance the comprehension of fundamental processed governing RNA dynamics and indeed the effectiveness of \emph{a posteriori} reweighting on RNA simulations has been established by several works \cite{borkar2016structure,krepl2017structural,podbevvsek2018structural,bottaro2018conformational,kooshapur2018structural}. The lack of a common strategy to increase the reliability of RNA force fields from first principles \cite{aytenfisu2017revised,tan2018rna,kuhrova2019improving}, however, makes it particularly appealing to resort to approaches that exploit discrepancies between MD simulations and experimental data to refine force fields. Here we review a recent study \cite{cesari2019fitting} that tackles this problem by proposing a likelihood minimisation scheme. Let us assume the system of interest is described by a force field $E_\text{ff}(X)$ and a corresponding Boltzmann probability distribution $P_0(X) \propto e^{-\beta E_\text{ff}(X)}$. For the same system, $M$ experimental data have also been collected. In this work, the authors seek for an optimised probability density
\begin{equation}
  P(X, \boldsymbol{\mu}) \propto P_0(X) e^{-\beta \sum_{i=0}^N\mu_i f_i(X)}
\end{equation} 
where now the force field is expressed by an expansion on a basis $f_i(X)$, where each function is associated to a weight $\mu_i$. We stress that weights $\boldsymbol{\mu}$ have not to be interpreted as the previously introduced weights $\boldsymbol{w}$, as the former are arbitrarily normalised and not interpreted as probability densities. Each of the $N$ terms helps in reproducing the $M$ experimental constraints, but, differently from MaxEnt methods, $f_i(X)$ do not represent the forward model connecting configuration $X$ to an experimental measurements. Rather, $f_i(X)$ can be generic functions and $N$ is sought in such a way that $N \ll M$.  While the analytic form of the basis functions is enforced at the beginning of the optimisation process, the value of the weights $\lambda_i$ needs to be determined through the minimisation of a function describing the discrepancy between the simulated and the experimental observables. We note that in this strategy the functional form of the force field is actively modified by introducing basis functions $f_i(X)$ that are not explicitly included in $E_\text{ff}(X)$. Alternatively, this can be seen as a way to associate non-zero weights to terms in the force field that have an effective null coupling constant. If functions  $f_i(X)$  are already part of the potential, instead, the corresponding weights $\mu_i$ amount for a refinement of the interaction strength. Supposing to be interested in $M$ observables $O_j^\text{calc}$,  the target for the minimisation takes the form of a regularised error function
\begin{equation}
T(\boldsymbol{\mu}) = T(\langle O_1^\text{calc}\rangle(\boldsymbol{\mu}), \ldots, \langle O_M ^\text{calc}\rangle(\boldsymbol{\mu}) ) + \theta|\boldsymbol{\mu}|^2 \qquad \theta \geq 0
\end{equation}
where the averages $\langle O_j\rangle(\boldsymbol{\mu})$ are computed in the refined ensemble
\begin{equation}
\langle O_j \rangle(\boldsymbol{\mu}) = \frac{1}{Z} \int dX\; O_j^\text{calc}(X) P(X, \boldsymbol{\mu}) 
\end{equation}
The error function is designed to enforce both equalities and inequalities, i.e. $\langle O_j^\text{calc} \rangle(\boldsymbol{\mu}) = O_j^\text{exp}$ and $\langle O^\text{calc}_j \rangle(\boldsymbol{\mu}) < O_j^\text{exp}$ and the optimal set of weights $\boldsymbol{\mu}$ is obtained as the one minimising $T$. The strength $\theta$ of the regularisation term is key to the strategy: for $\theta\to\infty$ the error function does not feel the contribution of the experimental constraints and thus the potential resulting from  the optimisation is just the original one $E_\text{ff}(X)$. Instead, for $\theta \to 0$ the deviations from the original potential are not restrained, and the prior information from the simulations is effectively ignored. Therefore, an optimal value of $\theta$ has to be determined or alternative statistical tools have to be applied to integrate out the variable. In this work the authors employ cross-validation to determine the optimal $\theta$, but this is a general issue common to many ensemble optimisation methods. We refer therefore to section \ref{sec:challenges} and references therein for further details on the choice of the optimal prior strength.  

The RNA systems chosen for this application were four tetranucleotides and two tetraloops, and NOEs (Nuclear Overhouser Effect) as well as scalar coupling NMR data were used in the force field refinement procedure. The basis functions were chosen to be sines and cosines acting on torsional angles, while the employed guess force field was  Amber ff99bsc0 + $\chi_\text{OL3}$ with the OPC water model \cite{perez2007refinement,zgarbova2011refinement,cornell1995second,izadi2014building}.

\section{Matching time-dependent and time-resolved data}\label{sec:time}
\begin{figure}[tbp!]
\begin{center}
\includegraphics[width=\textwidth]{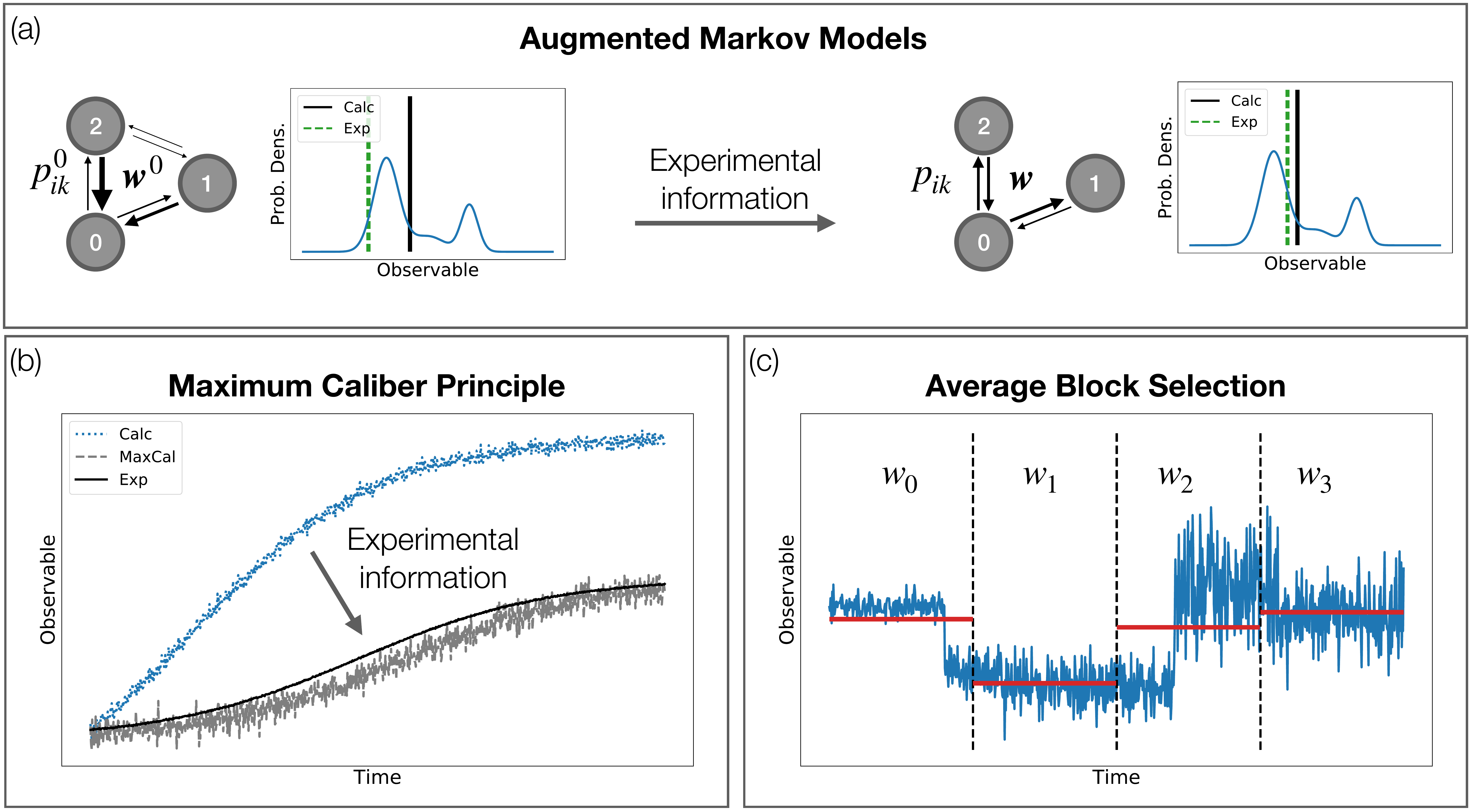}
\end{center}
\caption{Graphical summary of methods implemented to recover time-dependent and time-resolved data. (a) Augmented Markov Models employ experimental information to refine a guess Markov State Model and consequently better estimate kinetic observables of interests; (b) Maximum Caliber, in its experiment-biased formulation, is employed to bias the sampling to match some time-resolved experimental quantities; (c) Average block selection is used to assign weights to sub-trajectories and better estimate time-dependent experimental data.}
\label{fig:time_dep}
\end{figure}

When dealing with systems at equilibrium, the average of an observable $\langle O \rangle$ is consistent with many possible distributions of the same observable $p(O)$. As we discussed in section \ref{sec:reweighting}, MaxEnt, MaxPars and MaxPrior principles can help discriminate, among the several possibilities, which one is the optimal distribution. Suppose, however, we can access a good description of some relevant conformations of a molecule (by experimental evidences or previous sampling) and we are now interested in understanding the dynamics of interconversion between them. From an MD perspective, one can hope to see interconversions happening by running several trajectories starting from the collected relevant configurations. If extensive sampling is achieved and enough state transitions are captured, one can for example use the trajectories to build a Markov State Model for the system, and obtain the desired information about the inter-state dynamics \cite{pande2010everything, bowman2013introduction, husic2018markov}. However, full sampling of the conformational dynamics might be challenging to achieve, especially when the desired motions happen in the $\mu$s timescale or above. Moreover, as discussed in detail in section \ref{sec:FF}, limitations in the force field employed in the simulations might lead to conclusions which disagree with key experimental data, even in the case of excellent sampling. This may be the case, for example, for NMR spin relaxation experiments: spin relaxation rates are sensitive to both short- and long-range dynamics of a protein, occurring from the picosecond to the nanosecond timescale \cite{palmer2004nmr}. For dynamical observables like NMR spin relaxation rates, methods developed to increase the consistency between computed and experimental values of static observables might be of only limited help and different strategies are needed. In this section we describe some recent efforts in this direction and how it is possible to use simulations to match experimental knowledge on \emph{time-dependent} observables (i.e. observables that depend on time because of fast processes happening in the system and cannot therefore be expressed by means of Eq. \eqref{eq:ensemble_average}, e.g. NOEs with spin diffusion \cite{vasile2019determination}) and \emph{time-resolved} ones (i.e. observables that are at equilibrium locally in time, but have been monitored for long timescales and can therefore be expressed as in Eq. \eqref{eq:ensemble_average} by including a time label, e.g. time-resolved SAXS). The following sections will focus on applications concerning Maximum Entropy and Likelihood estimation applied to Markov State Models \cite{prinz2011markov,schwantes2014perspective} (section \ref{sec:time:maxlike}), the principle of Maximum Caliber \cite{jaynes1980minimum} (section \ref{sec:time:maxcal}) and average block selection \cite{salvi2016multi} (section \ref{sec:time:dyn_avg}) (Fig.~\ref{fig:time_dep}).

\subsection{Maximum Entropy and Likelihood in dynamical systems}\label{sec:time:maxlike}
In a kinetic model, the equilibrium distribution $\boldsymbol{w}$ can be determined from a transition probability matrix $T(\tau)$, i.e. a matrix whose elements $p_{ik}$ provide the probabilities for the system to be found in state $k$ at time $t+\tau$, given it was in the state $i$ at time $t$. $T(\tau)$, together with the structures of the states among which transitions happen, is usually known as a \emph{Markov State Model} (MSM) \cite{prinz2011markov,bowman2013introduction}.  In this section, we will discuss a method to refine a MSM by adding static (not time-dependent or time-resolved) experimental information. This framework, that balances data coming from simulations and experimental averages, is called Augmented Markov Model (AMM) \cite{olsson2017combining}. 

The construction of an AMM starts from the definition of an initial MSM, $T^0(\tau)$, with transition probabilities $p_{ik}^0$. The MSM is typically built by following a multi-step procedure \cite{prinz2011markov,bowman2013introduction}: (i) relevant features (e.g. dihedral angles, contact maps etc.) of the system are selected. The dynamics of these features needs to play a key role in the conformational transitions under consideration, so their identification might not be trivial. This task can become easier with the help of automatic selection tools \cite{bowman2009progress,sultan2014automatic}; (ii) the features are used to generate a lower-dimensional representation of the system dynamics \cite{perez2013identification, schwantes2013improvements}; (iii) the lower dimensionality \cite{noe2015kinetic} allows one to cluster similar configurations in the dimensionally reduced space. The $N$ clusters will define the states of the model; (iv) finally, statistical tools \cite{prinz2011markov} are employed to estimate the transition matrix $T^0(\tau)$. Among other possibilities, the transition probability matrix can be obtained by maximisation of the likelihood \cite{prinz2011markov,bowman2013introduction}:
\begin{equation}\label{eq:MSM_likelihood}
L(T^0) \propto \prod_{i,k}(p^0_{ik})^{c_{ik}},
\end{equation}where $c_{ik}$ is the number of transitions occurring between state $i$ and $k$, which are obtained by explicitly counting the transitions between states in the simulation \cite{scherer2015pyemma,harrigan2017msmbuilder}. The resulting transition matrix has to satisfy further criteria, e.g. it has to be row- or column-stochastic, and detailed balance can be enforced explicitly. Maximisation of Eq.~\eqref{eq:MSM_likelihood} is equivalent to minimising the negative log-likelihood:
\begin{align}\label{eq:S_MSM_likelihood}
S(T^0) = -\sum_{i,k}^N c_{ik}\log p^0_{ik},
\end{align}
We note that while Eq.~\ref{eq:S_MSM_likelihood} resembles an entropy term, but it cannot represent a proper one, 
as the counts $c_{ik}$ enter Eq. \ref{eq:MSM_likelihood} un-normalised. However, $S(T^0)$ can be used in a similar way as a regularising prior when including experimental information, as we describe further below. The equilibrium distribution of the system, usually referred to as the stationary probability distribution of the MSM, is obtained from the transition probability matrix by solving the eigenvalue equation $T^0(\tau)\boldsymbol{w}^0 = \boldsymbol{w}^0$. For more information about Markov State Models, we refer the reader to some excellent reviews on the subject \cite{pande2010everything,prinz2011markov,bowman2013introduction,husic2018markov,noe2019special}.

Given the initial Markov State Model $T^0(\tau)$, the framework of Augmented Markov Models employs the MaxEnt principle and likelihood maximisation to build an optimised MSM, $T(\tau)$. Let us assume we can access both the expectation values of $M$ experimental observables and an MSM built on simulation data, with the reference stationary distribution $\boldsymbol{w}_0$. To obtain the optimal distribution $\boldsymbol{w}$, one can apply MaxEnt principle, as reported in Eq. \eqref{eq:MaxEnt_2}, which bridges the model distribution with the experimental one by means of the provided experimental averages. Rather than estimating the weights corresponding to the optimised distribution $\boldsymbol{w}$, the MaxEnt formulation allows one to optimise the set of Lagrange multipliers $\boldsymbol{\lambda}$ (Eq.~\eqref{eq:MaxEnt_2}). As usual, the Lagrange multipliers are obtained by enforcing constraints on the experimental averages. In order to account for the statistical errors in both sampling and experiments, one can introduce the so-called \emph{Augmented Markov Model likelihood}, assuming Gaussian errors:
\begin{equation}\label{eq:AMM_likelihood}
L(\boldsymbol{w},T)\propto \left(\prod_{i,k}(p_{ik})^{c_{ik}} \right) e^{-\frac{1}{2}\chi^2(\boldsymbol{w})}. 
\end{equation}The term in the parentheses is the MSM likelihood, Eq. \eqref{eq:MSM_likelihood}, while the second term incorporates the Gaussian error model proposed for the observables. The corresponding negative log-likelihood is thus proportional to:
\begin{align}\label{eq:AMM}
T_\text{AMM}(\boldsymbol{w},T) = \chi^2(\boldsymbol{w}) - \theta S(T),
\end{align}
where $S(T)$ is the entropy-like term introduced in Eq. \eqref{eq:S_MSM_likelihood}. Note that $\boldsymbol{w}$ is uniquely determined by $T$ by solving the eigenvalue equation $T(\tau)\boldsymbol{w}=\boldsymbol{w}$, so Eq. \eqref{eq:AMM} is effectively a minimisation problem for the matrix elements $p_{ik}$ alone. We stress that we have introduced Eq.~\eqref{eq:AMM} only to show the similarity with the reweighting methods for equilibrium ensembles, Eq. \eqref{eq:MaxEnt_reg} and \eqref{eq:MaxPars}, and that it does not directly represent the way AMMs are practically implemented. Rather, one obtains an AMM that optimally balances between the information from simulation and the experimental data (Fig. \ref{fig:time_dep}a) by employing fixed-point iteration algorithm to maximise Eq. \ref{eq:AMM_likelihood} with respect to the unknowns $p_{ik}$ and the weights $\boldsymbol{w}$ (see the supplementary information of Ref. \cite{olsson2017combining} for further details). Olsson et al. \cite{olsson2017combining} applied this method to two 1~ms simulations of ubiquitin. They built AMMs using NMR scalar and residual dipolar couplings, i.e. static data, and their results were compared against NMR relaxation dispersion data. The results of such experiments can either be calculated directly from a long molecular dynamics simulation \cite{xue2012microsecond,lindorff2016picosecond} or indirectly from a MSM \cite{olsson2016mechanistic}. The authors found that the AMM that had been optimized against the experimental data was in overall better agreement this indpendent data compared to the raw MSM. This is a strong indication of a non-trivial fact: reweighting an MSM with respect to experimental equilibrium observables helps increase the reliability in the prediction of time-dependent data. 

\subsection{Maximum Caliber}\label{sec:time:maxcal}
The principle of Maximum Caliber (MaxCal) can be regarded as a generalisation of MaxEnt, and enables one to compute the probabilities associated to dynamical pathways, rather than probabilities (weights) associated with equilibrium states. To infer pathway probabilities, MaxCal principle maximises a \emph{path entropy} \cite{presse2013principles} defined over all the possible pathways, constrained to reproduce a given dynamical observable. The path entropy is defined as:
\begin{equation}\label{eq:path_entropy}
S[p_0(\gamma)] = -\sum_\gamma p_0(\gamma) \log p_0(\gamma),
\end{equation}where $p_0(\gamma)$ is the probability that the system follows the structural path $\gamma$. Each path is considered as a collection of configurations, $\gamma = \{X_0^{\gamma}, \ldots, X_T^{\gamma}\}$, labelled by a discrete time index $t=0,\ldots,T $. Maximisation of Eq. \eqref{eq:path_entropy}, together with $M$ constraints of the form:
\begin{equation}
g^j[p_0(\gamma)] = 0 \quad j = 1,\ldots,M
\end{equation}
yields the optimal distribution $p(\gamma)$ over pathways. Just like in the MaxEnt case, constraints are usually enforced by Lagrange multipliers, and typically they concern the calculated dynamical observables $O^\text{calc}(X_t^\gamma)$ and corresponding experimental values at the same time point $t$, $O_t^\text{exp}$:
\begin{equation}\label{eq:pMaxCal_constraint1}
\sum_{\gamma} p_0(\gamma)O^\text{calc}(X_t^\gamma) -  O^\text{exp}_t = 0
\end{equation}The normalisation of the probability distribution of pathways is also enforced:
\begin{equation}\label{eq:pMaxCal_constraint2}
\sum_\gamma p_0(\gamma) -1 = 0.
\end{equation}
If it is possible to introduce an estimate for the time-dependent error $\sigma_t$, then the notion of $\chi^2$ can be extended to a time-dependent form as well, $\chi^2=\chi^2_t$. In this way, by analogy with MaxEnt, MaxCal principle can also be expressed in a regularised fashion as
\begin{equation}\label{eq:MaxCal_funct}
T_\text{MaxCal}[p_0(\gamma)] = \chi^2_t[p_0(\gamma)]  - \theta_t S[p_0(\gamma)]
\end{equation}where the parameter $\theta$ of Eq. \eqref{eq:MaxEnt_reg} acquires a dependence on time. All the terms in Eq. \eqref{eq:MaxCal_funct} are functionals of the probability density in the space of pathways: as such, the minimisation of $T_\text{MaxCal}[p_0(\gamma)]$ would require a search in path space, which is knowingly a hard task \cite{bolhuis2002transition,eastman2001simulation,pan2008finding,weinan2002string}. To our knowledge, indeed, MaxCal principle has never been used in the acceptation of Eq. \eqref{eq:MaxCal_funct}. Instead, the principle of Maximum Caliber can be employed to run restrained MD simulations, where restraints are provided by time-resolved experimental data (Fig. \ref{fig:time_dep}b).

Note, when comparing simulations with time-resolved experiments, a single simulation cannot be used to generate an ensemble as in the MaxEnt method (section \ref{sec:reweighting}), due to the non-equilibrium conditions. Therefore, replicas are needed to describe ensemble development over time (see also section \ref{sec:on-the-fly}). Capelli et al. \cite{capelli2018implementation} introduced such replica-averaging implementation of MaxCal, so ensembles can be determined at each time $t$. In this formulation, Eq. \eqref{eq:pMaxCal_constraint1} and \eqref{eq:pMaxCal_constraint2} are employed to constrain the minimisation of the path entropy, together with the path probability density normalisation and two more equations. The first one constrains the system's diffusion constant $D$:
\begin{equation}\label{eq:pMaxCal_constraint3}
\frac{1}{2\Delta t} \sum_\gamma p_0(\gamma) [X^{\gamma}_{t+1} - X^{\gamma}_t]^2 - D = 0,
\end{equation}while the other constrains the standard deviation $\sigma_t$ of the observable of interest at time $t$, obtained by averaging it over all the $N$ replicas
\begin{equation}\label{eq:pMaxCal_constraint4}
\sum_{\gamma^i} p_0(\gamma^i) \left( \langle O^\text{calc}(X^{\gamma^i}_t) \rangle - O_t^\text{exp} \right)^2 - \sigma_{t}^2  = \sum_{\gamma^i} p_0(\gamma^i)\xi_t^2 - \sigma_{t}^2 = 0,
\end{equation}
where $\sum_{\gamma^i}$ is used to specify that the sum is carried out over all the possible pathways of all replicas and the average $\langle \cdot \rangle$ is computed over replicas. The expression of the path entropy, together with all the constraints (in this particular case, Eq. \eqref{eq:pMaxCal_constraint1}, \eqref{eq:pMaxCal_constraint2}, \eqref{eq:pMaxCal_constraint3} and \eqref{eq:pMaxCal_constraint4}) imposed via Lagrange multipliers is called \emph{caliber}. The result of caliber maximisation provides the optimal distribution of pathways:
\begin{equation}\label{eq:pMaxCal_distribution}
p(\gamma) = \frac{1}{Z}\exp\left[ -\sum_{t,i} (\nu^i_t \left[X_{t+1}^{\gamma} - X_t^{\gamma}\right]^2+\lambda^i_t O(X_t^{i,\gamma}) + \mu_{t}^i \xi_t^2) \right],
\end{equation}
where $Z$ is a normalisation factor, $\nu_t^i$, $\lambda_t^i$ and $\mu_t^i$ are replica- and time-dependent Lagrange multipliers.  Despite the complicated expression of the optimal path probability distribution, it is possible to prove that, in the assumption where the system's dynamics is Brownian and by analogy with MaxEnt \cite{capelli2018implementation, roux2013statistical,cavalli2013molecular}, the MaxCal distribution in Eq. \eqref{eq:pMaxCal_distribution} can be sampled by adding a time-dependent, harmonic bias potential to the underlying force field:
\begin{equation}\label{eq:pMaxCal_potential}
E_\mathrm{exp}(\gamma^i, t) = Nk\left( \langle O^\text{calc}(X^{\gamma}_t) \rangle -O_t^\text{exp}  \right)^2,
\end{equation}
where the constant $k$ is used to tune the strength of the interaction. This approach was applied to the case of the second hairpin of  protein G B1 domain by employing synthetically generated time-resolved SAXS data \cite{capelli2018implementation}. \\

Rather than using MaxCal principle as an experiment-biasing technique, it is also possible to employ it for reweighting, in particular for the case of MSMs \cite{dixit2018caliber,dixit2018communication,bause2019microscopic}. Suppose we built a MSM from a set of trajectories and we want to predict how a given non-equilibrium observable $O_{ik}$ changes in the transition between state $i$ and state $k$. A possible example would be the change in a spectroscopic signal (e.g. F\"{o}rster resonance energy transfer or circular dichroism) when passing from a partially unfolded state $i$  to a helical state $k$ of a protein. The average, computed over many transitions, is provided by:
\begin{equation}\label{eq:cal:average}
\langle O^\mathrm{calc} \rangle(\boldsymbol{w}) = \sum_{i,k} w^0_{i} p_{ik}O^\mathrm{calc}_{ik},
\end{equation}where $\boldsymbol{w}^0$ is the stationary probability distribution of the MSM. Let us suppose we also possess experimental information of such observable, $O^\text{exp}$ and that the computed average does not match the expected result. In this case, we can use the MaxCal principle to select among the models with the correct average \cite{dixit2018communication}. To do so, first the entropy is built as:
\begin{equation}\label{eq:entropy_MSM}
S(T,\boldsymbol{w}) = - \sum_{i,k}w_i p_{ik}\log\left( \frac{p_{ik}}{p^0_{ik}}\right),
\end{equation}where $\boldsymbol{w}$ is the optimal stationary distribution and $p_{ik}$ are the  refined transition probabilities. Note that Eq. \eqref{eq:entropy_MSM} is a version of Eq. \eqref{eq:path_entropy} in the case of discrete pathways. The entropy has to be maximised together with the constraint that the function
\begin{equation}\label{eq:MSMe_c1}
g(\boldsymbol{w}) = \langle O^\mathrm{calc} \rangle(\boldsymbol{w}) - O^\text{exp}
\end{equation}
is zero, where $\langle O^\mathrm{calc} \rangle(\boldsymbol{w})$ is the average over transitions introduced in Eq. \eqref{eq:cal:average}. Moreover, $w_i$ and $p_{ik}$ are interdependent because of probability conservation, so three more constraints naturally emerge:
\begin{equation}\label{eq:MSMe_c2}
\sum_{i}w_i p_{ik}-w_k = 0 \quad \sum_{k}w_i p_{ik}-w_{i} = 0 \quad \sum_{i,k}w_i p_{ik}-1 = 0.
\end{equation}
The minimisation of the entropy in Eq. \eqref{eq:entropy_MSM}, with the constraints in Eq. \eqref{eq:MSMe_c1} and \eqref{eq:MSMe_c2}, can be carried out with the method of Lagrange multipliers. It yields:
\begin{equation}\label{eq:cal:opt_trans}
p_{ik} = \eta\phi_k\phi_i^{-1}M_{ik}(\lambda,p^0_{ik},O^\text{calc}_{ik})
%\qquad M_{ik} = p_{ik}\exp(-\lambda\, O_{ik}),
\end{equation}
where $M_{ik}(\lambda,p^0_{ik},O^\text{calc}_{ik})$ is a non-Hermitian matrix, $\boldsymbol{\phi}$ is its only right-eigenvector having only positive elements (whose existence and uniqueness is guaranteed by the Perron-Frobenius theorem), $\eta$ the corresponding eigenvalue and $\lambda$ is the Lagrange multiplier controlling the constraint in Eq. \eqref{eq:MSMe_c1}. The positivity of the elements of vector $\boldsymbol{\phi}$ is necessary to guarantee the positivity of the transition probabilities. Eq. \eqref{eq:cal:opt_trans} provides a new set of transition probabilities defining an optimal transition probability matrix $T(\tau)$.  Effectively, similarly to what happened in the case of AMMs, the introduction of experimental information actively modifies the transition probabilities between states, and interconversions can become more or less favourable depending on the cases. Finally, the optimal stationary distribution is obtained as:
\begin{equation}
w_i = \psi_i\phi_i,
\end{equation}where $\boldsymbol{\psi}$ is the left-eigenvector of matrix $M_{ik}(\lambda,p^0_{ik},O^\text{calc}_{ik}) $ corresponding to the eigenvalue $\eta$. Matrix $M_{ik}(\lambda,p^0_{ik},O^\text{calc}_{ik}) $ only depends on the Lagrange multiplier $\lambda$, which is used as a parameter, and on quantities that can be readily computed from the initial MSM, i.e. the transition probabilities and the observables of interest. Therefore, the optimal transition probabilities and equilibrium probability distribution can be obtained by single value decomposition of $M_{ik}(\lambda,p^0_{ik},O^\text{calc}_{ik}) $ for different values of $\lambda$: the optimal Lagrange multiplier will be the one for which the constraint in Eq. \eqref{eq:MSMe_c1}, estimated \emph{a posteriori} from the updated MSM, holds true.

In the original work, the method was tested on a toy model for  a growth factor activation pathway, showing how the incorporation of experimental information can help correcting the transition probabilities in a guess MSM. Despite the method, as illustrated here, does not incorporate experimental errors, it is possible to take them into account in a Bayesian fashion \cite{dixit2018communication}. 

\subsection{Average Block Selection}\label{sec:time:dyn_avg}
Time-dependent data can also be applied directly in the ensemble refinement by means of time-dependent average block selection. Suppose we have $N$ MD trajectories, started from different conformations, and a set of $M$ time-dependent experimental observables $O^\text{exp}_j$. As anticipated at the beginning of this section, by time-dependent we here mean observables that cannot be calculated using Eq. \eqref{eq:ensemble_average} because the values depend on the underlying dynamics as well as on the stationary distribution. To increase the accuracy of the MD predictions, given the experimental data, and preserve information about dynamical processes, Salvi et al. \cite{salvi2016multi} proposed to divide each trajectory into $B$ blocks representing subsequent time-windows. The total number of blocks is then $N\times B$, each one associated to a weight $w_b$. The block-weights are then optimised by minimising the residual sum:
\begin{equation}\label{eq:absurd}
\mathcal{R}^2(\boldsymbol{w}) = \sum_{j=1}^M \left( O^\text{exp}_j-\sum_{b=1}^{N\cdot B} w_bO_{bj}^\text{calc} \right)^2 
\end{equation}
where $O_{bj}^\text{calc}$ is the $j$-th observable computed from the $b$-th block. In later studies, the authors included errors in the expression by the usual $\chi^2(\boldsymbol{w})$ expression \cite{Salvi2019}.  Division in blocks allows to determine how much each block contributes to the time-dependent experimental signal. For example, blocks with null weights are excluded and can be interpreted as non-physical artefacts. Despite the similarity with other approaches used for equilibrium observables, this method shows two important differences: on the one hand, weights are not associated to a single structure of an ensemble, but rather to sub-trajectories (Fig. \ref{fig:time_dep}c). It is evident, then, that the method can be effective only in the case where the experimental timescales of interest can be sampled within a single trajectory. On the other hand the method as it stands is unregularised, i.e. it assumes that the optimal solution is obtained by minimising the sum of residuals without adding any restraint. This is a crucial difference with respect to MaxEnt, MaxPars and MaxPrior strategies: one would expect nonetheless the method to benefit from regularisation terms. The addition of such terms would be advisable in further applications, but it is not clear what kind of prior would be needed in this case. Because of the fact that Eq. \eqref{eq:absurd} associates weights to pathways and not conformations, it might be possible to explore a connection with \emph{a posteriori} reweighting using MaxCal principle. 

In its original formulation, the method (called ABSURD ---  Average Block Selection Using Relaxation Data) \cite{salvi2016multi} was designed to deal in particular with NMR spin relaxation rates, but its application can be extended to any experimental source sharing similar timescales in a straightforward way. ABSURD was applied to the C-terminal domain of the nucleoprotein of Sandai virus. A cumulative time of approximately 6.5 $\mu$s of MD was sampled, employing two different water models. While the reproduction of spin relaxation rates by MD simulations alone was imperfect, the ABSURD-optimised trajectories were found in much better agreement with relaxation data on a wide spectrum of timescales, even in the simplest case where a single experimental rate was employed in the optimisation. 

\section{Challenges}\label{sec:challenges}
In sections \ref{sec:reweighting}--\ref{sec:time} we focused on major efforts that have been done to merge information coming from experiments and simulations to refine conformational ensembles or improve existing force fields. Despite many important steps forward in the last decade, some major challenges remain. In this section we will point out some of these and try to draw a possible paths for solutions. In section \ref{sec:balance} we will focus on the problem of setting the parameter $\theta$ in a robust way. In section \ref{sec:rew_ff} we will discuss the option of combining force field corrections and reweighting for a given system. In section \ref{sec:kinetic} we discuss possibilities and obstacles to employ kinetic data to reweight equilibrium ensembles. Finally, section \ref{sec:new_generation} is dedicated to a discussion on the next generation of force fields.  

\subsection{Balance between simulations and experimental data}\label{sec:balance}
In the MaxEnt and MaxPars reweighting methods, and equivalent methods biasing on the fly, the prior knowledge coming from simulations is balanced against experimental data by tuning the parameter $\theta$ (Eqs. \eqref{eq:MaxEnt_reg}, \eqref{eq:MaxPars} and \eqref{eq:on-the-fly}). In principle, the balance could be exactly known if the effective uncertainties on the weights, force field, sampling, and on the both the calculated and experimental observables were known (as also discussed in section \ref{sec:exp_bias:maxprior}). However, the uncertainties on the weights are usually unknown and the experimental uncertainty stems from different sources, only some of which are typically known or easy to estimate. We can subdivide the sources to the experimental uncertainty in the following categories: (i) Statistical errors on experimental data; (ii) statistical errors in the calculation of average observables from a limited amount of structures (Eq. \eqref{eq:ensemble_average}); (iii) systematic errors on experimental data; (iv) inaccuracy of the forward model. For normally distributed errors, these add up to a total variance \cite{Hummer2015}. In the following we will shortly discuss each of them.

The statistical error on experimental data is usually estimated by repetitive measurements and counting statistics, and it is included in most approaches, e.g. as standard deviation in the $\chi^2$. Over- or underestimated errors may be detected by visual good fits having $\chi^2_r$ much lower or higher than unity.

The statistical error on the mean of the observables calculated from the ensemble can easily be taken into account \cite{Bonomi2017}. This is important when the ensemble is small, which is typically the case for MaxPars methods and experiment-biased approaches with replicas.

Systematic errors in experimental data are generally difficult to take into account, as their magnitude and nature are usually unknown and may be system specific (e.g. incorrect buffer subtraction in SAXS \cite{Shevchuk2017}). Shevchuk and Hub \cite{shevchuk2017bayesian} treated the systematic errors with Bayesian statistics as a \emph{nuisance} parameter, i.e. an unknown (and uninteresting) parameter that should be determined together with the model parameters. Bonomi et al. \cite{Bonomi2016} discussed the more general case of outliers, and showed that an approach that combines reweighting and experiment-biased force fields is more robust against outliers than other related methods. Similar methods are also discussed by K\"ofinger et al. \cite{kofinger2019efficient}.

Inaccuracies in the forward model are likewise highly non-trivial to estimate. While negligible in some cases, they are the dominant error source in others, e.g. NMR chemical shifts, where they can be orders of magnitude greater than the statistical error. SAXS forward models share similar problems \cite{Cordeiro2017}. The source of the model inaccuracy might also come from neglecting an intrinsic time-dependency of the observable (e.g. spin diffusion or dynamic effects in the estimation of NOEs \cite{vasile2019determination}). 

In summary, as long as errors in the force field, systematic errors in the experimental data, and errors on the forward model continue to be challenging, if not impossible, to estimate accurately, the parameter $\theta$ remains necessary as an effective scaling of the total error $\sigma$. A key challenge is therefore to determine it. A simple way is to tune $\theta$ until $\chi^2_r$ reaches unity (Eq. \eqref{eq:rchi2}). This is, however, not generally a valid approach   \cite{Andrae2010a}. First, the number of degrees of freedom $\nu$ is ill-defined when reweighting is concerned. Conventionally, $\nu$ is estimated as $M-k$, where $M$ is the number of data points and $k$ is the number of fitted parameters. In MaxEnt, for example, $k$ is given by the number of Lagrange multipliers, $k=M$, therefore $\nu$ is effectively zero and $\chi^2_r$ has consequently no meaning. Second, although the expectation value of $\chi^2_r$ is one, point estimates are typically different from unity and $\chi^2_r$-distributed. Therefore, this method may provide incorrect balance between data and simulation. Another strategy to determine $\theta$ is to plot $S(\boldsymbol{w})$ against $\chi^2(\boldsymbol{w})$ and look for an \emph{elbow} in the curve \cite{bottaro2018conformational,kofinger2019efficient}. Indeed, if plotted with double-logarithmic axis, $\chi^2(S)$ often becomes an L-shaped curve, and the optimal $\theta$ can be estimated by finding the kink of the curve \cite{Hansen2000_2}. A third strategy is cross-validation, i.e. fitting the optimal ensemble to a separate set of data that has not been used in the analysis \cite{Lindorff-Larsen2005,Bottaro2018saxs, cesari2019fitting, Chen2019,crehuet2019bayesian}. The reweighting or experiment-biased simulation may then be done for several values of $\theta$ to monitor when the goodness of fit to the unused data starts to decrease. While intuitively appealing, the practical implementation may be difficult. First, the data needs to be divided into independent subsets, which may sometimes be difficult for highly correlated or interdependent data. Second, as different sources of data may report on very different aspects of the system, they may in practice not be useful for cross validation \cite{crehuet2019bayesian}. A fourth strategy for determining $\theta$ is strictly Bayesian, and $\theta$ is in that context treated as a nuisance parameter. The posterior probabilities at each value of $\theta$ are calculated and used to find the most probable value of $\theta$ \cite{Larsen2018}, or to integrate out $\theta$ completely \cite{Hummer2015}.

\subsection{Interplay between reweighting and force field corrections}\label{sec:rew_ff}
In section \ref{sec:compare} we mentioned that reweighting might fail in reproducing the experimental averages when the prior (force field) used is too inaccurate, because relevant states are either poorly sampled or not sampled at all. A pragmatic, and potentially transferable, solution to this problem would be to identify the parameters in the force field responsible for the incorrect behaviour and modify them slightly in order to move closer to the expected averages. Such rescaling approaches have proved succesful for specific systems and force fields, e.g. for adjusting the protein-protein interaction strength in the coarse-grained Martini force field \cite{Stark2013,Javanainen2017} or protein-water interactions for simulations of disordered proteins \cite{best2014balanced}. A subsequent reweighting of the obtained trajectories might correct for inconsistencies that are not related to the identified and re-scaled parameter, and could therefore lead to better consistency with the experimental data.

Applying such approaches, however, raises some relevant questions. For example, to what extent should one correct the force field? Considerable modifications to a force field would typically require the need to iteratively re-sample the system under consideration, an exercise that might become computationally expensive. Moreover, after any substantial reparametrisation the force field should be benchmarked against other data and ab initio calculations, as done in the original parametrisation of the force field. Therefore, should one perhaps modify the force field just enough to allow for reweighting? And, more generally, should force field corrections and reweighting be employed together at all? Standard force field reparameterization effectively corresponds to reweighting a conformational ensemble, but the extent is limited by the functional form of the energy function, and the simultaneous consideration of data on other systems. A concurrent application of reweighting and force field optimization could violate the Bayesian principle of having well-defined prior and likelihood in the reweighting process, as the re-scaled prior used for reweigting has already been adjusted against experimental data. The data is, so to say, used twice in such protocol. We do not have an answer to the aforementioned questions, nonetheless we believe they provide important points of discussion for future applications of reweighting, and that Bayesian methods for force field parameterization would make it easier to merge these two different strategies. 

\subsection{Using kinetic data to reweight equilibrium ensembles} \label{sec:kinetic}
As discussed in section \ref{sec:time:maxlike}, Augmented Markov Models are a framework based on Maximum Entropy and Maximum Likelihood to include experimental information in a Markov State Model. Olsson et al. \cite{olsson2017combining} used AMMs to predict NMR relaxation dispersion data, by employing only static experimental data to reweight the model. This fact leads to intriguing questions which, to our knowledge, have been largely unanswered in literature. As a matter of principle, it should be possible  to directly employ kinetic data for ensemble reweighting, just like it is done with the ABSURD method \cite{salvi2016multi}. What would happen then to equilibrium observables? Would the amount of provided information be enough to increase the accuracy of simulated averages with respect to experimental equilibrium quantities? What would be the optimal framework to test this hypothesis? While the first question has no clear answer yet, concerning the latter we believe that a suitable framework would be the one of Markov State Models, because of their intrinsic ability to encode kinetic information \cite{olsson2017combining,dixit2018caliber}. We note however that MSMs, by construction, ignore the fastest dynamical timescales of the system, which can however be important for some types of experimental measurements. Therefore, a careful choice of kinetic data is recommended. 

\subsection{A new generation of force fields}\label{sec:new_generation}
Force field parametrisation has been historically guided by a combination of chemical and biological intuition, ab initio quantum calculation of small molecules and trial and error approaches \cite{zhu2012recent}. This approach has been proven to be extremely successful \cite{hospital2015molecular} but, as the ever growing amount of experimental information calls for force fields which are easily improvable once new data are collected, the complexity and the amount of expertise required in this procedure have made it somewhat impractical. We discussed some advancements in section \ref{sec:ff_opt}, but none of the applications we reviewed provided an ultimate solution. Indeed, suggested improvements are usually small adjustments (e.g. modifications in the backbone dihedral angles terms \cite{li2010nmr,li2011iterative}), though in some cases more extensive changes have been introduced using such fitting to experiments \cite{wang2012systematic,wang2014building,wang2017building,robustelli2018developing}. Therefore, in this section we want to discuss some principles on which a new generation of force fields could be build.

One of the main challenges in force field development lies in how to include new data, which is potentially conflicting with old information, and use it to improve the force field after it has already been parametrised. 
Automatic reparametrisation would make it extremely easy to modify and update force field once such new experimental data become available. The development of a Bayesian formalism \cite{norgaard2008experimental} and later the more systematic ForceBalance \cite{wang2012systematic,wang2014building,wang2017building} framework, for example, goes exactly in this direction and shows that it is possible in principle to approach to the problem in an automated fashion. It would be also important to assign some level of \emph{trust} to force field parameters, such that it can be assessed to what degree new data should alter their values. At the same time, a fully Bayesian approach would involve distribution of force fields, rather than point estimates, and thus parallel simulations could be used to integrate out force field uncertainty. We also stress that such developments would ideally be carried concurrently with the construction of worldwide accessible and curated databases of experimental and simulation data. In order to include new data in a more automatic fashion, there have to be data quality checks and consensus on experimental and forward model errors, as this all affects how much the new data should be able to alter the existing parametrisation (in case of inconsistency). At the same time, it is also important to keep in mind that such models should ideally capture well-understood physical effects, and that lack of agreement with experiments might indicate important effects that are missing from the functional form or parameter combining rules \cite{hagler2019force}.

The molecular mechanics force fields were developed as a classical parametrisation of the Born-Oppenheimer energy surface that balances computational speed and precision and the choices of functional forms of the force field terms has always been guided by chemical intuition. This has led to families of force fields with differences in the parametrisation \cite{Guvench2008} and to the proliferation of big sets of parameters needed to accommodate empirical choices (see, for example, the discussion on atom types in Refs. \cite{mobley2018escaping,zanette2018toward}). We expect the next generation of force fields to be more flexible with respect to specific choices of parametrisation: more specifically, tools are required to define not only the force field parameters, but also the functional shapes of each term in a data-driven fashion. These developments go hand in hand with a robust estimation of both statistical and systematic errors: the two sources of errors need to be fully decoupled and it should be clear when poor estimates are due to poor training datasets. 

Some of these advancements have already been applied to the SMIRNOFF99Frosst force field and the Open Force Field Toolkit \cite{mobley2018escaping}, developed by the Open Force Field Initiative \cite{openff}. For now, SMIRNOFF99Frosst has been tested only on a wide set of pharmaceutically relevant small molecules, but it represents nonetheless an important step towards a new generation of force fields.

\section{Conclusions}\label{sec:conclusions}
Much research in the field of structural biology and molecular biophysics requires one to integrate several heterogeneous sources of data, coming both from diverse experimental techniques and simulations. Experiments are employed to characterise thermodynamic and kinetic quantities of the system under study and simulations aid the interpretation of or complement these results thanks to their high spatial and time resolutions. However, technological \cite{piana2013atomic,lindorff2016picosecond,voelz2012slow,bowman2010atomistic} and theoretical \cite{Maximova2016,camilloni2018advanced} advancements in the field of molecular mechanics have highlighted the existence of discrepancies between simulations and experiments \cite{rauscher2015structural,henriques2015molecular,robustelli2018developing,nerenberg2018new,bernardi2015enhanced,Cordeiro2017}. In this review, we have approached this issue through a specific perspective: inconsistencies between computational and experimental results carry information that can be systematically extracted and exploited to improve our understanding of biochemical entities and more general biophysical models. To pursue this perspective, we focused on the cornerstone ideas that have guided the optimisation of system-specific conformational ensembles (sections \ref{sec:reweighting} and \ref{sec:on-the-fly}) or general purpose force fields (section \ref{sec:FF}) against experimental information. We provided a summary of alternative strategies and discussed their strengths and shortcomings depending on the level of trust one places in the force field used to carry out the simulations. For each method, we critically assessed how much a realistic scenario deviates from the ideal version of the approach, examining the challenges that arise when the systems grow in size and complexity. We stressed that many sources of data can be employed and different strategies are needed depending on whether observables can or cannot be represented as a linear combination of a given forward model applied to the single structures in the ensemble (see Eq. \eqref{eq:ensemble_average} and section \ref{sec:time}). Despite the differences in the technical implementation, however, we argued that most of the approaches presented here share a common framework, rooted in the minimisation of the functional
\begin{equation}
T = \chi^2 - \theta R,
\end{equation}
where $\chi^2$ is the negative log-likelihood in the case of normally distributed errors, $\theta$ is a hyperparameter that balances between simulations and experimental data and $R$ is a regularisation term. Depending on the method, $R$ assumes different functional shapes: in MaxEnt, it becomes a cross-entropy term, Eq. \eqref{eq:MaxEnt_reg}; in MaxPars, it reduces a parsimony term, i.e. the negative number of conformations with non-zero associated weight, Eq. \eqref{eq:MaxPars}; in MaxPrior, the regularisation term is provided by the log-prior, Eq. \eqref{eq:MaxPrior}; in AMMs $R$ represents the logarithm of the MSM likelihood, Eq. \eqref{eq:AMM}; in MaxCal the regularisation term takes the form of a path-entropy, Eq. \eqref{eq:MaxCal_funct}; finally, average block selection is unregularised and so $R=0$, Eq. \eqref{eq:absurd}. Therefore, the different philosophies that distinguish among the several approaches are encoded in the choice of the regularisation term. 

As a final remark, we devoted section \ref{sec:challenges} to the challenges that the community is facing and we anticipate some open questions that we believe will capture experts' attention in the incoming years. 

\section*{Acknowledgements}
We acknowledge support by a grant from the Lundbeck Foundation to the BRAINSTRUC structural biology initiative, the NordForsk Nordic Neutron Science Programme, the Carlsberg Foundation, a grant from the Velux Foundations and a Hallas-M{\o}ller Stipend from the Novo Nordisk Foundation. We would like to thank members of the Linderstr{\o}m-Lang Centre for Protein Science for numerous discussions on these topics, and thank Giovanni Bussi, Ramon Crehuet, Clemens Kauffmann and Simon Olsson for insightful comments on the manuscript. 

\bibliographystyle{ieeetr}
\bibliography{references}
\end{document}